\documentclass{pasj01}
\usepackage[utf8]{inputenc}
\usepackage[english]{babel}
\usepackage{graphics}
\usepackage{pgfplotstable}
\usepackage{appendix}
\usepackage{savesym}
\usepackage{natbib}
\usepackage{xspace}
\usepackage{booktabs}
\usepackage{longtable}
\usepackage{morefloats}
\usepackage{blindtext}
\usepackage{placeins}
\usepackage{lineno}
\savesymbol{iint}
\savesymbol{leftroot}
\savesymbol{idotsint}
\savesymbol{uproot}
\savesymbol{iiint}
\savesymbol{iiiint}
\usepackage{amsmath}
\restoresymbol{AMS}{iint}
\restoresymbol{AMS}{iiint}
\restoresymbol{AMS}{iiiint}
\restoresymbol{AMS}{leftroot}
\restoresymbol{AMS}{uproot}
\restoresymbol{AMS}{idotsint}
\usepackage{makecell}
\usepackage[shortlabels]{enumitem}
\usepackage[caption=false]{subfig}
\pgfplotsset{compat=1.17}
\draft 
\Received{$\langle$reception date$\rangle$}
\Accepted{$\langle$acception date$\rangle$}
\Published{$\langle$publication date$\rangle$}

\begin{document}
\title{Gamma-Ray Bursts, Supernovae Ia and Baryon Acoustic Oscillations: a binned cosmological analysis}

\author{ M. G. Dainotti$^{1,2}$}
\altaffiltext{1}{National Astronomical Observatory of Japan, Division of Science, Mitaka, 2-chome;}
\altaffiltext{2}{Space Science Institute, Boulder, Colorado}
\email{maria.dainotti@nao.ac.jp;}

\author{G. Sarracino$^{3,4}$}
\altaffiltext{3}{Dipartimento di Fisica, ``E. Pancini''
Universit\`{a} ``Federico II'' di Napoli, 
Compl. Univ. Monte S. Angelo Ed. G, Via Cinthia, I-80126
Napoli (Italy);}
\altaffiltext{4}{INFN Sez. di Napoli, Compl. Univ. 
Monte S. Angelo Ed. G, Via Cinthia, I-80126 Napoli (Italy);}

\author{S. Capozziello$^{3,4,5}$}
\altaffiltext{5}{Scuola Superiore Meridionale, Università di Napoli Federico II
Largo San Marcellino 10, 80138 Napoli (Italy);}


\maketitle

\begin{abstract}

Cosmological probes at any redshift are necessary to reconstruct consistently the cosmic history. Studying properly the tension on the Hubble constant, $H_0$, obtained by Supernovae Type Ia (SNe Ia) and the Planck measurements of the Cosmic Microwave Background Radiation would require complete samples of distance indicators at any epoch. Gamma-Ray Bursts (GRBs) are necessary for the aforementioned task because of their huge luminosity that allows us to extend the cosmic ladder at very high redshifts. However, using GRBs alone as standard candles is challenging because their luminosity varies widely. To this end, we choose a reliable correlation for GRBs with a very small intrinsic scatter: the so-called fundamental plane correlation for GRB afterglows corrected for selection biases and redshift evolution. We choose a well-defined sample: the platinum sample, composed of 50 Long GRBs. To further constrain cosmological parameters, we use Baryon Acoustic Oscillations (BAOs) given their reliability as standard rulers. Thus, we have applied GRBs, SNe Ia, and BAOs in a binned analysis in redshifts so that GRBs' contribution is fully included in the last redshift bin, which reaches $z=5$.
We use the fundamental plane correlation together with SNe Ia and BAOs, to constrain $H_0$ and the density matter today, $\Omega_{M}$.
This methodology allows us to assess the role of GRBs combined with SNe Ia and BAOs. We have obtained results for $H_0$ and $\Omega_{M}$ using GRBs+ SNe Ia+BAOs with better precision than the SNe Ia alone for every bin, thus confirming the beneficial role of BAOs and GRBs added together. In addition, consistent results between GRBs+ SNe Ia +BAOs are obtained when compared with the SNe Ia +BAOs, showing the importance of GRBs since the distance ladder is extended up to $z=5$ with a similar precision obtained with other probes without including the GRBs.
\end{abstract}

\section{Introduction} \label{Introduction}
Supernovae Type Ia (SNe Ia) are considered reliable standard candles because of their well established and nearly uniform luminosities (corresponding to a mean absolute magnitude around $M\sim-19.5$, \citealt{Carroll}), but the maximum redshift observed until now for a SN Ia is $z=2.26$ \citep{Rodney2015}. This means that they cannot be used to probe the early stages of the universe. The so-called $H_0$-tension between the Cosmic Microwave Background (CMB) radiation measurement ($H_0=67.4 \pm 0.5$, \citealt{Planck2018}), and the ones related to the SNe Ia and the Cepheids ($H_0=74.03 \pm 1.42$, \citealt{Riess2019})  has been observed. An investigation of this tension would allow possibly to bridge the gap between the late and early Universe. One possible explanation for this tension could be that this is actually a signature for new physics that goes beyond General Relativity \citep{Dainotti2021a, Benetti, Capozziello, Dainotti2022a, Spallicci2022}.

To this end, we need astrophysical observables detectable at high redshift, like Gamma-Ray Bursts (GRBs). 
GRBs are the most luminous panchromatic short-lived phenomena \citep{Paczynski1986}. Their luminosity allows us to observe them up to $z=9.4$ \citep{Cucchiara}, further than many other luminous cosmological objects, like SNe Ia and quasars. The latter have been very recently observed at $z=7.642$ \citep{Mortlock2011, Wang2021}. 
Similarly to quasars, the application of GRBs as cosmological probes is very challenging \cite{Dainotti2022b} because their physical mechanism is still a puzzle to be solved.
This panorama is further complicated by the irregularity of the GRB features and light-curves (LCs). The GRB scenario was complex and mysterious 50 years ago when they were first discovered and, although we have cast some light on both their origins and physical mechanisms, many questions remain not completely answered. 
Indeed, there are various proposed models regarding their birth, such as the coalescence of two compact objects, like  neutron stars (NSs) and black holes (BHs, \cite{Lattimer, Eichler1989, Metzger2014, Rowlinson, Rea2015, Piro2017, Ciolfi2017, Stratta, Fujibayashi2018, Piro2019, Tanaka2020}), or the core-collapse of very massive stars \citep{Narayan1992, Woosley1993, MacFadyen1999}. These scenarios consider  BHs, NSs, or highly magnetized fast spinning NSs (magnetars, \cite{Ai2018}) as the engine of GRB emission. In the former case, the engine is created after the merging of two compact objects and its subsequent explosion, while, in the latter one, it is the compact object left as a remnant after the collapse of the massive star.

In this puzzling scenario, further light can be shed if we can link the progenitor's nature with the GRB phenomenology. In this regard, in the past 30 years, there have been many efforts done by the scientific community to classify and standardize GRBs according to their observable features. 
GRB LCs usually consist of a prompt phase followed by the so-called afterglow.
The prompt is generally observed in the hard X-rays and $\gamma$-rays. The afterglow phase is a multi-wavelength emission, usually observed in X-rays, optical, and also radio bandwidths, after the prompt emission \citep{Perley, Morsony, Li2015, O'Brien, Sakamoto2007}. 

One traditional, widely accepted GRB classification categorizes them according to the duration of their prompt emission: 
the long GRBs have $T_{90}\ge 2$ s \footnote{$T_{90}$ is the time during the prompt emission in which the GRB releases from $5\%$ to $95\%$ of its total measured photons.}, while the short GRBs are characterized by $T_{90} \leq 2$ s \citep{Mazets, Kouveliotou, Bromberg2013, Lu2014}. Nevertheless, this classification takes into account only the energy and the time interval of the prompt emission, while a second classification has been proposed based on their different progenitors \citep{Zhang2009, Kann, Berger, Li, Fraija}, and on how they can be deduced by the GRBs' observable physical features (i.e., host galaxy, prompt duration, presence of a natal kick), although the discussion about the natal kick is still controversial \citep{Hagai}. According to this classification, GRBs born by the merging of two compact objects are called Type I GRBs, while the ones originated after the collapse of a massive star are denominated Type II GRBs. 

The Neil Gehrels Swift Observatory (hereafter Swift, \citealt{Gehrels}) can detect the GRB plateau emission in the X-ray wavelength \citep{O'Brien, Sakamoto2007, Evans}, usually lasting  $10^2-10^5$ s and followed by a decaying phase which can be fitted in the majority of cases by a power-law (PL).

Several scenarios can describe the plateau emission: two of the most accredited ones deal with an external shock between the relativistic ejecta and the surrounding interstellar medium powered by energy injections from the central engine, \citep{Paczynski1993, Meszaros1997, Wijers1997, Piran1998, Sari1998, Zou2005, Zhang2006, Granot2012, Gao2013, Yi2013} or with the luminosity generated by the spin-down of a magnetar \citep{Thompson, Dai1998, Metzger2008, Metzger2017, Metzger2018, Dall'Osso2011, Zhang2011, Bernardini, Giacomazzo2013, Gompertz, Gompertz2014, Stratta, Surorov}. Another model for the plateau is based on a two-component emission: the first one linked to a prior ejecta of energy happened before the main burst, while the second to the main outflow itself \citep{Yamazaki2009}. 
The plateau emission may be originated from the fallback material into the BH \citep{Kumar2008a, Kumar2008b, Cannizzo2009, Cannizzo2011, Kisaka, Gao2016, Gibson}.
Correlations involving the plateau have been deeply studied and investigated by \citet{Dainotti2008, Dainotti2011a, Dainotti2011b, Dainotti2013a, Dainotti2015a, Dainotti2017a, Dainotti2020b, Dainotti2021a, Liang,  Xu2012, Del Vecchio, Shun-Kun2018, Tang2019, Zhao2019, Srinivasaragavan, Wen2020}. These correlations have been further applied as cosmological tools \citep{Cardone,Postnikov,Dainotti2013b,Dainotti2022d}. 
One of these relations is the so-called Dainotti relation, which links the rest frame time at the end of the plateau, $T^{*}_X$, to its corresponding luminosity, $L_X$ \citep{Dainotti2008, Dainotti2010, Dainotti2011a, Dainotti2017a}. This correlation finds its natural explanation within the magnetar framework \citep{Zhang2001,Rowlinson,Rea2015,Liang2018,Stratta}. Within this scenario, the most important implication is that the energy supply of the plateau remains roughly constant during its duration. This correlation has also been extended in optical and it has been interpreted within the same magnetar scenario \citep{Dainotti2020b}. \cite{Dainotti2011b, Dainotti2015b} discovered the existence of the correlation between the prompt emission's peak luminosity, $L_{peak}$ and  $L_X$. Thus, the combination of the $L_{peak}-L_X$ and the $L_X$-$T^{*}_X$ has led to a three-dimensional relation between $L_X$-$T^{*}_X$-$L_{peak}$ \citep{Dainotti2016, Dainotti2017b, Dainotti2020a, Dainotti2021c}. 
An extension of this correlation in the optical range has been recently presented \cite{Dainotti2022c}.
Such an extension is named the GRB fundamental plane, or the 3D Dainotti relation. It is important to note that for using GRB correlations properly a segregation among classes is necessary \citep{Dainotti2010}, especially in view of the diversity of the density rate evolution and luminosity functions between Short and Long GRBs, for details see \cite{Dainotti2021b}. For a review on GRB correlations, their selection biases and application as cosmological tools see \cite{Dainotti2017c, Dainotti2018a, Dainotti2018b}.

To infer the Hubble constant, $H_0$, and the density of the matter content in the universe today, $\Omega_M$, through the fundamental plane correlation, we select a GRB subsample with very well-defined properties, so that the dispersion of the correlation becomes as small as possible. An ideal candidate for this purpose is the platinum sample (hereafter PLAT sample) introduced in \citet{Dainotti2020a}, and described in section \ref{sample selection}. To further enhance the reliability of this correlation as a cosmological tool we need to guarantee that we have appropriately considered the selection biases and redshift evolutionary effects. As a result of the correction of these effects through the reliable \citet{Efron}, hereafter EP, statistical method, we have shown that we obtain one of the smallest intrinsic scatter, $\sigma_{int}$, in the literature for any existing GRB correlation, especially for those involving plateau features. A detailed discussion about the EP method and its applications to GRBs and to the fundamental plane correlation is shown in the following papers: \cite{Lloyd2000, Dainotti2013a, Dainotti2015a,  Dainotti2020a, Petrosian2015}. 
As shown in \cite{Dainotti2020a}, using the fundamental plane relation related to the PLAT sample corrected for the selection biases and redshift evolution leads to a central value of $\sigma_{int}$ 35\% smaller than the one derived from the observed fundamental plane relation for the same sample that it is not corrected for selection biases and redshift evolution. 

Together with the notion of standard candles, another definition should be considered in a multi-probe analysis of cosmological models. Indeed, the inclusion of the so-called standard rulers is necessary for a deep understanding of the universe's structure. The standard rulers are astrophysical phenomena whose dimensions or characteristic scales are known and can be used to measure the distances with respect to the redshift values \citep{Ntelis2018}. In this case, Baryon Acoustic Oscillations (BAOs) are particularly important. These variations in the density of the detectable baryonic matter manifest as a single spike in the correlation function given by the interference of baryonic and dark matter perturbations: at the epoch of recombination, the typical scales of BAOs are around $150$ Mpc \citep{Eisenstein2005}. This peak of the correlation function at $150$ Mpc is explained by \cite{Eisenstein2005} in the following way: at the time of the recombination, the expanding shell of the acoustic wave is around the peak mentioned above, after which the dark matter and baryonic perturbations start the formation of large structures.

The layout of the paper is as follows: in section \ref{sample selection} we detail the three data samples composed of GRBs, SNe Ia, and BAOs. In section \ref{Methodology SNe} we present the analysis performed on the SNe Ia data. In section \ref{Methodology BAOs} we show how we use BAOs as constraints for cosmological parameters. In section \ref{Methodology GRBs} we describe the best fit of the fundamental plane relation, both neglecting and including the correction for selection biases and evolutionary effects. In section \ref{Methodology Cosmology} we detail our approach to obtain $H_0$ and $\Omega_M$ derived from the binned analysis. In section \ref{Results} we show the results using SNe Ia only, SNe Ia+BAOs, and SNe Ia+BAOs+GRBs. For the GRB samples we consider both the cases with and without evolution in each bin and in the total sample. Lastly, in section \ref{Conclusions}, we draw and discuss our conclusions.

\section{Sample Selection for GRB\lowercase{s}, SN\lowercase{e}, and BAO\lowercase{s}} \label{sample selection}

To select the GRB sample, we adopted the same criteria applied in \citet{Srinivasaragavan} and \citet{Dainotti2020a}. In this work we consider the PLAT sample only, defined in \citet{Dainotti2020a}, for which the plateau must fulfill the following morphological criteria: \newline
1) its beginning should have at least five data points; \newline
2) its inclination should be smaller than $41^{\circ}$ which means that the plateau should not be too steep. This criteria follows the choice for the Gold Sample already defined and used in \citet{Dainotti2016,Dainotti2017a}, where it has been found that the angles are characterized by a Gaussian distribution from which the outliers beyond 1-$\sigma$ level have been cut from that Gaussian, for details see the figure 10 showing the histograms of the angles in Appendix B in \cite{Dainotti2017a};
\newline
3) both its beginning and its end times should not fall within observational gaps so that we have a more precise measurement of its duration;\newline 
4) its total duration should be at least $500$ s; \newline
5) it should not show observed flares or bumps within the time duration of the plateau. Given the criterium 3 this should assure us that there is a low probability of flares during this phase.

Starting from an initial sample of 222 GRBs presenting a plateau, we extract 50 GRBs according to the above-mentioned criteria that will be part of our PLAT sample, which have redshifts ranging from $z=0.055$ up to $z=5$. This sample is a subsample of the Gold Sample, which has already been used successfully for the fundamental plane correlation, with even more strict morphological criteria. This is the main reason why we have chosen this sample. We use the \citet{W07} fitting model employing the BAT + XRT LCs from the Swift repository. \footnote{http://www.swift.ac.uk/burst\texttt{\_}analyser}

For the SNe Ia set, we employ the Pantheon Sample (Hereafter PS, \citealt{Scolnic}) that has been built by collecting $1048$ SNe Ia from different surveys. The PS ranges from $z=0.01$ to $z=2.26$. 
The PS, together with the other samples, has been divided into 5 bins in order of increasing redshift. 

\begin{figure}
    \centering
    \includegraphics[width=0.7\hsize,height=0.6\textwidth,angle=0,clip]{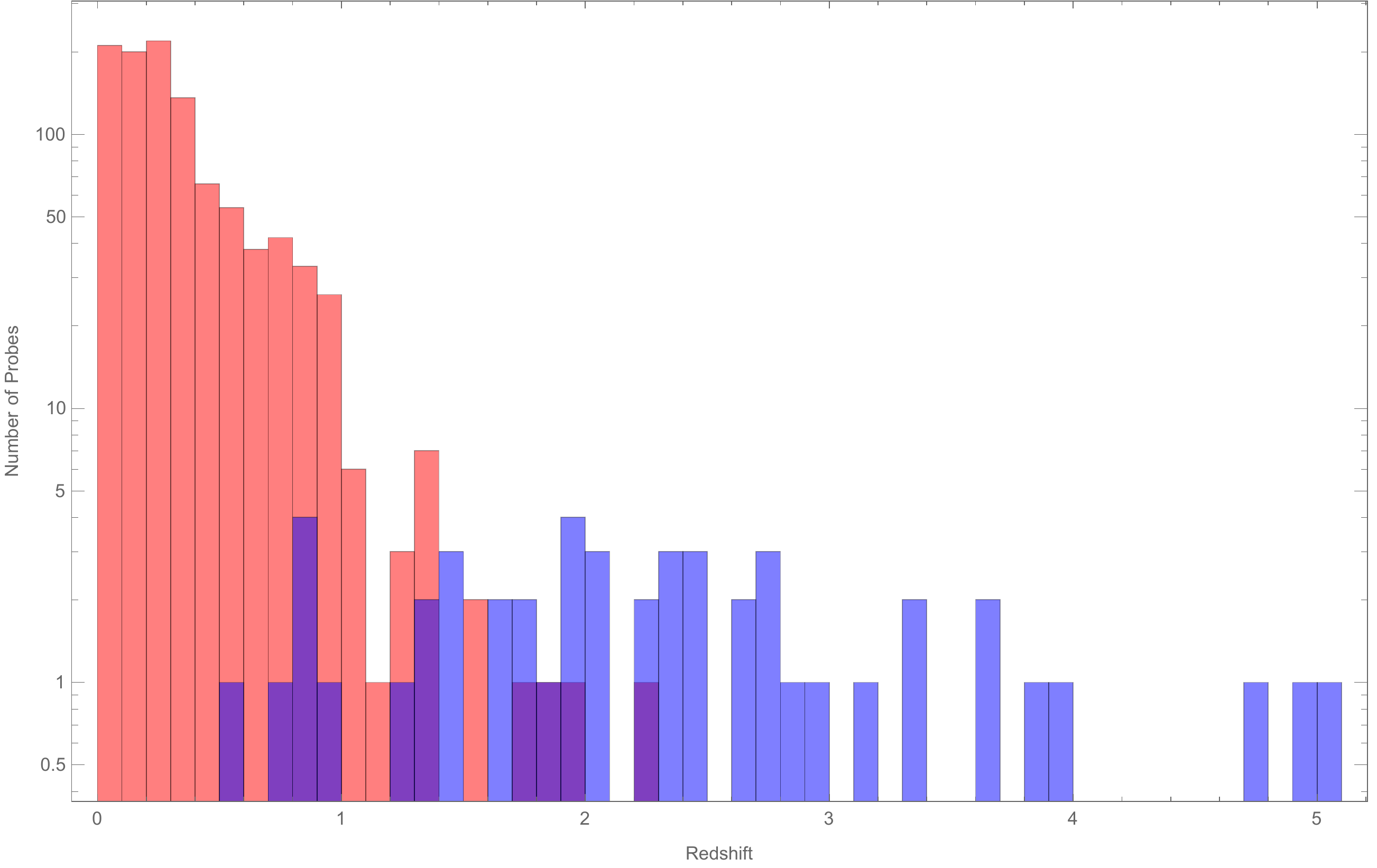}
    
    \caption{\textbf{The overlapped histograms of the redshift of both SNe Ia (red) and GRBs (blue), considering a logarithmic scale for the number of probes}}
    \label{RedshiftHistogram}
\end{figure}

Regarding the BAO related data, we used 16 measurements \citep{Beutler, Blake, Ross, du Mas des Bourboux, Alam2021}, whose nature will be detailed in section \ref{Methodology BAOs}. We here note that the redshift range of these measurements goes from $z=0.106$ to $z=2.36$.
The differences in the redshift distributions of both SNe Ia (red) and GRBs (blue) are shown in figure \ref{RedshiftHistogram}. In particular, we can visualize how much our PLAT sample extends beyond the redshift of the SNe Ia.


\section{Methodology for the SN\lowercase{e} Ia analysis} \label{Methodology SNe}
To compute the best fit value for $H_0$ using SNe Ia, we start from the observed distance modulus, $\mu_{obs, SNe}$, collected from the SNe Ia detections gathered in \citet{Scolnic}, and we compute its difference from the theoretical $\mu_{th}$. The latter is linked to the luminosity distance by the following equation: 

\begin{equation}
\mu_{th}=5\cdot \log_{10} d_L(z,\Omega_{M}, H_0, w) +25,
\label{modulus}
\end{equation}
where $d_L$ is the luminosity distance in Mpc, defined in equations (\ref{flatdistanceluminosity}) and (\ref{SNeDL}) in section \ref{Methodology Cosmology}.

\noindent The observed distance modulus of the SNe Ia is defined as:

\begin{equation}
\label{modulus SNE}
\mu_{obs, SNe}=m_{B}-M+\alpha x_{1} - \beta c + Q, 
\end{equation}

\noindent where $c$ is the color, $x_{1}$ is the stretch parameter, $M$ is the absolute magnitude in the B-band of a reference SN with $x_{1}=0$ and $c=0$,  $m_{B}$ is the apparent magnitude in the B-band, and $Q$ is a constant. 

\noindent In particular, $x_{1}$ is the factor that parametrizes the LCs time scale \citep{Perlmutter}: a basic template LC is used and then stretched to match the observations, so that we have an actual standard candle. Furthermore, the color used here is the color index $B-V$ computed at the maximum of the SNe Ia LC in blue and visual bands, once all the corrections due to the reddening brought by the host galaxy dust are accounted for \citep{Phillips}. 
There are different models that can be applied to a sample $x_{1}$ and $c$ for a specific SNe Ia population, in particular the G10 \citep{Guy} and the C11 \citep{Chotard} models. The former attributes more weight to the chromatic variation of the \textbf{SNe Ia} population and concludes that the interpretation of SNe Ia colors can be ascribed to a color dispersion and a variation of SNe Ia spectral features. The latter is mainly based on achromatic variations and its results show that there is no explicit redshift evolution of the coefficient $\beta$ present in equation (\ref{modulus SNE}).

Because there is no clear indication on which of these two models is the most suitable, considering the mean of C11 and G10 bias corrections is a plausible assumption. This assumption has indeed been taken following \cite{Scolnic,Dainotti2021a} and \cite{Dainotti2022a}. This average builds up the systematic covariance matrix, $C_{sys}$, which added to the statistical uncertainties, $D_{stat}$, constitutes the full covariance matrix denoted with $\mathcal{C}$, whose inverse will be used in the SNe Ia likelihood. 
Regarding the division in submatrices, we consider the mean of the two models and the full covariance matrix. We also choose to consider the mean of the two values of $\mu_{obs}$ given by the two models presented above. We implement the analysis by binning the data similarly to the approach used in \citet{Kazantzidis} and \citet{Dainotti2021a}. The differences of our computations from the ones performed by \citet{Kazantzidis} are the following: 
\begin{itemize}
\item
we divide our sample into more bins (5 vs. 4), so that we have all the GRBs of the PLAT sample gathered inside bin 5,
\item we use the average of the matrices of C11 and G10,
\item
we compute $H_0$ and $\Omega_M$ rather than $\mathcal{M}$, that is instead analyzed in \citet{Kazantzidis}
\item
 we use GRBs and BAOs combined with SNe Ia.
\end{itemize}
Besides, our investigation is different from the ones in \citet{Dainotti2021a, Dainotti2022a}, because we here divide the PS in 5 redshift bins instead of 3, 4, 20, and 40 bins, and we include GRBs and BAOs in our computations.

\section{The BAO\lowercase{s} applied as cosmological constraints} \label{Methodology BAOs}
The BAOs are caused by the propagation of acoustic waves with ions (namely, the baryons) into the relativistic plasma medium before the recombination epoch. The last scattering surface is
at redshift $z_d\approx1059$, where the decoupling of photons occur, and manifests as a maximum of the correlation function that describes the galaxies distribution at the scale of the sound horizon $r_s(z_d)$ \citep{Sharov2018}.

The BAO based measurements used in our work are taken from the 6dF Galaxy Survey \citep{Beutler}, the WiggleZ Dark Energy Survey \citep{Blake}, and the latest Sloan Digital Sky Survey (SDSS) data release (in particular, the Baryon
Oscillation Spectroscopic Survey (BOSS), and the extended Baryon
Oscillation Spectroscopic Survey (eBOSS)) \citep{Ross, du Mas des Bourboux, Alam2021} counting a total of 16 BAO related measurements in a redshift interval of $0.10 \leq z \leq 2.33$.

The quantities detected via the BAOs considered in our analysis are different from one another, contrary to the data at our disposal for GRBs and SNe Ia. Indeed, in \cite{Beutler, Ross} and for the eBOSS  emission line galaxies (ELGs, \citealt{Alam2021}) the dilation scale, $D_V(z)$, is provided, as dependent from the product of the radial dilation and the square of the transverse dilation:

\begin{equation}
    D_V(z)=\biggr[{D_M}^2(z)\frac{cz}{H(z)} \biggr]^\frac{1}{3},
    \label{eq_dilationscale}
\end{equation}

\noindent where $H(z)$ is the Hubble Parameter computed at a given redshift and $D_M(z)$ is the comoving angular diameter distance at the same redshift. In \cite{Blake}, instead, a different quantity, $A(z)$, linked to equation \ref{eq_dilationscale}, is provided, defined as follows:

\begin{equation}
    A(z)=\frac{100D_V(z)\sqrt{\Omega_m h^2}}{cz},
    \label{eq_dz}
\end{equation}

where $h=H_0/(100 km \hspace{1ex} s^{-1} Mpc^{-1})$. Finally, the remaining measurements \citep{Alam2021, du Mas des Bourboux}, are related to the comoving angular diameter distance $D_M(z)$ and the Hubble distance $D_H(z)$, defined as:

\begin{equation}
    D_H(z)=\frac{c}{H(z)}.
        \label{dH}
\end{equation}

\textbf{Many of the quantities measured by BAOs and described above are reported rescaled by $r_s(z_d)$. To give an estimate of this parameter in order to match our models with the observations,} we apply the approximated formula shown in \textbf{\cite{Aubourg,Sharov2016}}, namely:

\begin{equation}
    r_s(z_d)=\frac{55.154 \cdot e^{[-72.3(\Omega_{\nu}h^{2}+0.0006)^2]}}{(\Omega_{M}h^{2})^{0.25351}(\Omega_{b}h^{2})^{0.12807}}Mpc,
    \label{eq_rsfiducialtrue}
\end{equation}

\noindent where $\Omega_{b}$ is the baryonic density in the universe, which has been set to the value measured by \cite{Planck2018} ($\Omega_{b}\cdot h^2=0.02237$), and $\Omega_{\nu}$ is the neutrino density in the universe, that again has been fixed to the value provided by the $\Lambda$CDM ($\Omega_{\nu} \cdot h^2=0.00064$). The loglikelihood function for BAOs can be written in the form:

\begin{equation}
    \mathcal{L}_{BAO}=\log  \frac{1}{\sqrt{2\pi |\mathcal{C}_{BAO}|}}-\frac{1}{2}\Delta d^T \times \mathcal{C}_{BAO}^{-1} \times \Delta d,
    \label{eq_chi2_BAOs}
\end{equation}

\begin{table}
\centering

\scalebox{0.92}{%
\begin{tabular}{|c|c|c|c|c|c|c|}
\hline
number &$z$BAO & Measured Quantity & Value & Covariance & Data Group& Reference \\\hline
1 & 0.106 & $D_v(z)$ & $456 \pm 27$ Mpc &  no & 6DF Galaxy Survey & \cite{Beutler} \\\hline
2 & 0.44 & $A(z)$ & $0.474$ &  yes (with 3 and 4) & WiggleZ & \cite{Blake} \\\hline
3 & 0.60 & $A(z)$ & $0.442$ & yes (with 2 and 4) & WiggleZ & \cite{Blake} \\\hline
4 & 0.73 & $A(z)$ & $0.424$ &  yes (with 2 and 3)  & WiggleZ & \cite{Blake} \\\hline
5 & 0.15 & $D_v(z)\frac{r_{s,fid}}{r_s}$ & $664 \pm 25$ Mpc &  no & SDSS DR7 & \cite{Ross} \\\hline
6 & 2.33 & $\frac{D_H(z)}{r_s}$ & $8.99 \pm 0.19$ &  no & eBOSS (Ly$\alpha$) & \cite{du Mas des Bourboux} \\\hline
7 & 2.33 & $\frac{D_M(z)}{r_s}$ & $37.5 \pm 1.1$ &  no & eBOSS (Ly$\alpha$) & \cite{du Mas des Bourboux} \\\hline
8 & 0.38 & $\frac{D_M(z)}{r_s}$ & $10.23$ &  yes (with 9,10,11) & BOSS DR12 & \cite{Alam2021} \\\hline
9 & 0.38 & $\frac{D_H(z)}{r_s}$ & $25.00$ &  yes (with 8,10,11) & BOSS DR12 & \cite{Alam2021} \\\hline
10 & 0.51 & $\frac{D_M(z)}{r_s}$ & $13.36$ &  yes (with 8,9,11) & BOSS DR12 & \cite{Alam2021}\\\hline
11 &  0.51 & $\frac{D_H(z)}{r_s}$ & $22.33$ &   yes (with 8,9,10) & BOSS DR12 & \cite{Alam2021} \\\hline
12 & 0.698 & $\frac{D_M(z)}{r_s}$ & $17.86$ &  yes (with 13) & eBOSS LRG & \cite{Alam2021} \\\hline
13 & 0.698 & $\frac{D_H(z)}{r_s}$ & $19.33$ &  yes (with 12) & eBOSS LRG & \cite{Alam2021} \\\hline
14 & 0.65 & $\frac{D_v(z)}{r_s}$ & $18.33 \pm 0.60$ & no & eBOSS ELG & \cite{Alam2021} \\\hline
15 & 1.48 & $\frac{D_M(z)}{r_s}$ & $30.69$ &  yes (with 16) & eBOSS QUASAR & \cite{Alam2021} \\\hline
16 & 1.48 & $\frac{D_H(z)}{r_s}$ & $13.26$ &  yes (with 15) & eBOSS QUASAR & \cite{Alam2021} \\\hline
\end{tabular}}
\caption{The table summarizes the 16 BAO measurements used in our analysis. In the second column we note the redshift of the measurement; in the third the particular quantity measured as detailed in the main text; in the fourth column the value of such a quantity (with the error in the case of no covariances); in the fifth column we see if there are non-zero covariances with other measurements and, if yes, with which one; in the sixth column we report to which data group they belong to and, lastly, in the seventh column we report the reference papers from where we have taken the measurements. For the fifth data point, $r_{d,fid}=148.69$ Mpc. Apart from data points 1 and 5, all the other quantities are dimensionless.}
\label{Table:BAO}
\end{table}

\noindent where $\Delta d=d^{obs}_z(z_i)-d^{th}_z(z_i)$ correspond to the different quantities presented above and $\mathcal{C}_{BAO}^{-1}$ is the full inverse covariance matrix for the BAOs data points which includes both statistical uncertainties on the diagonal and off-diagonal terms indicating a correlation between different measurements. 
Each point of the covariance matrix is given by the different sources of BAO data used in our work, including the off-diagonal correlation terms. We stress that these 16 data points have been derived by hundred of thousands of observations related to galaxies, clusters of galaxies, quasars and Lyman $\alpha$ lines \citep{Alam2021}. A description of each BAO data point is shown in Table \ref{Table:BAO}.

\section{Methodology for GRBs: the application of the GRB fundamental plane as a cosmological tool} \label{Methodology GRBs}
To compute the best fitting parameters of the fundamental plane relation as well as the best fit value of the cosmological parameters $H_0$ and $\Omega_M$, we employ the \citet{Dago05} and \citet{ Reichart} Bayesian methods. From now on and in all the paper, scatters on the values are always expressed in 1 $ \sigma$. The luminosities are described by the following formula:

\begin{equation}
    log_{10}L_X=4\cdot \pi \cdot d_L(z_{GRB},\Omega_M, H_0)^2 \cdot F_X \cdot K,
    \label{eq_luminosityplateau}
\end{equation}

\noindent where $d_L(z_{GRB},\Omega_M, H_0)$ is the luminosity distance computed for a given GRB at redshift $z_{GRB}$, $F_X$ is the flux measured in the X-Ray band $(erg \hspace{1ex} cm^{-2} s^{-1})$ at the end of the plateau emission over a $1$ s interval, and $K$ is the $K$-correction considering the cosmological expansion \citep{Bloom}. This correction is calculated in the following way \citep{Bloom}:

\begin{equation}
K=\frac{\int_{E_\mathrm{{min}}/(1+z)}^{E_\mathrm{{max}}/(1+z)} \Phi(E)dE}{\int_{E_\mathrm{{min}}}^{E_\mathrm{{max}}} \Phi(E) dE} \, ,
\label{kcorrection}
\end{equation}
where $\Phi(E)$ is the functional form for the spectrum dependent on the energy $E$, for which we assume either a PL for the plateau phase and a cut-off power law for the prompt emission, while $E_{min}$ and $E_{max}$ are the minimum and maximum energies in which the spectrum is integrated, respectively. For the cases in which the spectrum is a PL, the $K$-correction becomes $\textit{K}=(1+z)^{(\beta-1)}$ \citep{Dainotti2017a}, where $\beta$ is the X-ray spectral index of the plateau phase, derived considering \cite{Evans}.

In equation (\ref{eq_luminosityplateau}) $d_L(z_{GRB},\Omega_M, H_0)$ depends also on the cosmological parameters, which for the fundamental plane fitting are fixed at the fiducial values of the $\Lambda$CDM model: $H_0= 70 \hspace{1ex} km \hspace{1ex} s^{-1} Mpc^{-1}$ and $\Omega_M=0.3$. The same formula has been used for $L_{peak}$ as well, but considering the peak flux.

\noindent The 3D Dainotti correlation is described by the following equation:
\begin{equation} 
\log_{10} (L_X /1 erg)= c + a \cdot \log_{10} (T^{*}_{X}/1 s)  + b \cdot \log_{10} (L_{peak}/1 erg)
\label{jetted},
\end{equation} 
\noindent where $c$ is a normalization, while $a$ and $b$ are the coefficients linked to $\log_{10} T^{*}_{X}$ and $\log_{10} L_{peak}$, respectively. The parameters $a$, $b$ and $c$ are all dimensionless and are computed with the D'Agostini fitting. We here report the results we have obtained in \citet{Dainotti2020a} by using the PLAT sample: $a = -0.86 \pm 0.13$,
$b =0.56 \pm 0.12$, $c = 21.8 \pm 6.3$. The intrinsic scatter, $\sigma_{int}$ (also dimensionless), computed for this plane is $\sigma_{int}=0.34 \pm 0.04$.
We also show the results found in \cite{Dainotti2020a} when the plane has been corrected for selection biases through the EP method. The fundamental plane, after this correction, takes the form:

\begin{equation}
\begin{split}
\log_{10} (L'_X/1 erg)-\alpha \log_{10}(z+1) =\\
a_{ev} \cdot(\log_{10} (T'^{*}_{X}/1s)- \beta \log_{10} (z+1))+ b_{ev} \cdot(\log_{10}(L'_{peak}/1 erg)- \gamma \log_{10}(z+1))+C_{ev},
\label{plane evolution}
\end{split}
\end{equation}
\vspace{0.25 cm}

\noindent where $\alpha$, $\beta$ and $\gamma$ are the coefficients related to the evolution taken from \cite{Dainotti2017b} and implemented in \cite{Dainotti2020a} for a wide range of GRB subclasses. This correction enters as a further factor since $L'$, which denotes the variable de-evolved is equal to $L_{X}'=L_{X}/(z+1)^{\alpha}$, and the same applies to the other variables: $L_{peak}'=L_{peak}/(z+1)^{\gamma}$, and  $T'^{*}_{X}=T^{*}_{X}/(z+1)^{\beta}$. Taking into account these effects, the values of the best fit parameters become: $a_{ev}=-0.90 \pm 0.16$, $b_{ev}=-0.50 \pm 0.16$, $C_{ev} = 25.6 \pm 8.2$, while the intrinsic scatter is $\sigma_{int}=0.22 \pm 0.10$, which has a central value that is $35 \%$ smaller than the one computed from the fundamental plane not corrected for selection biases. Again, all the coefficients involved in this correlation are dimensionless, like for the case without evolution. The results obtained by the Reichart method are compatible within 1 $\sigma$ both with and without the corrections due to the EP method.

\begin{figure}

\includegraphics[width=0.33\hsize,height=0.3\textwidth,angle=0,clip]{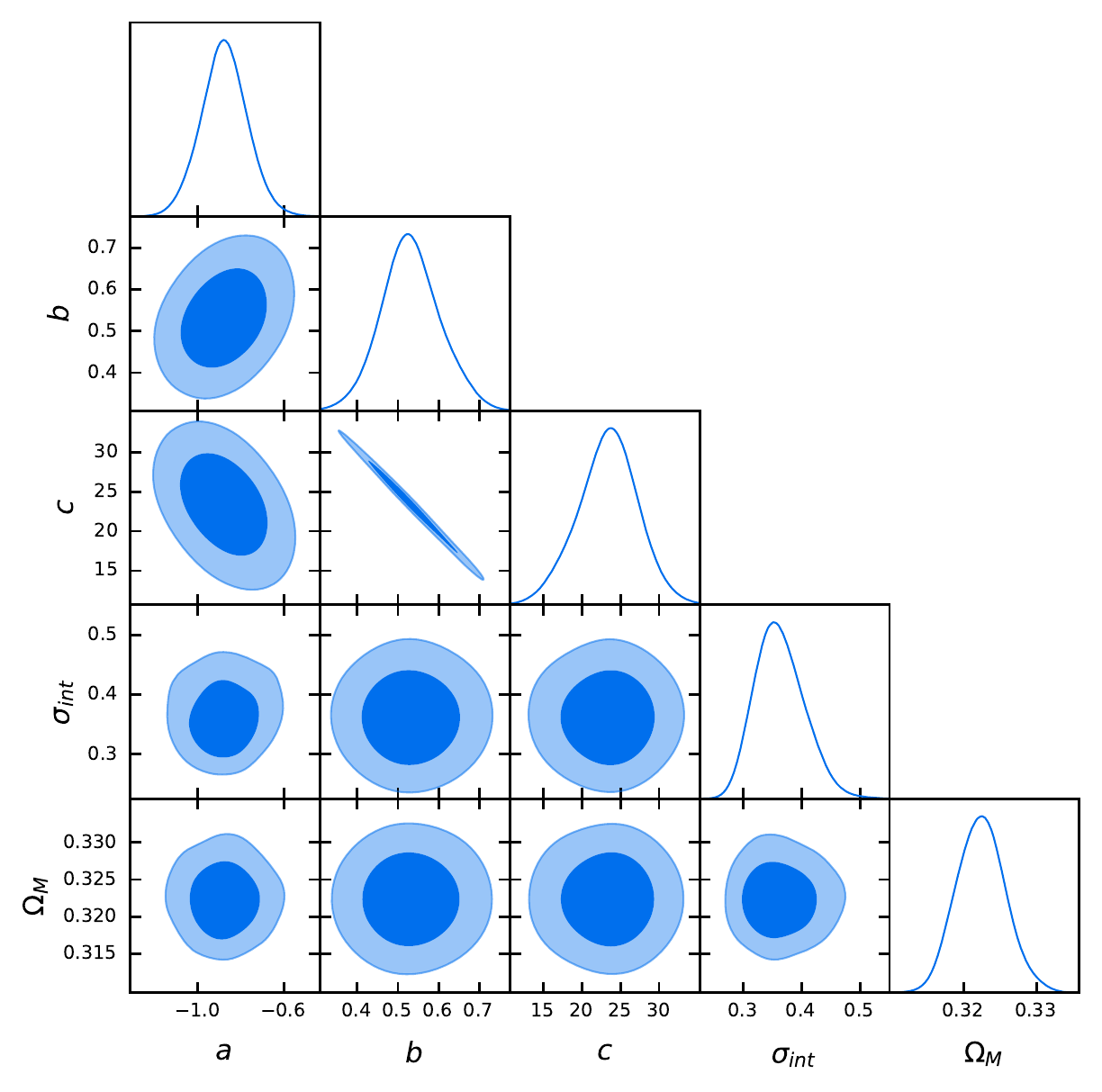}
\includegraphics[width=0.33\hsize,height=0.3\textwidth,angle=0,clip]{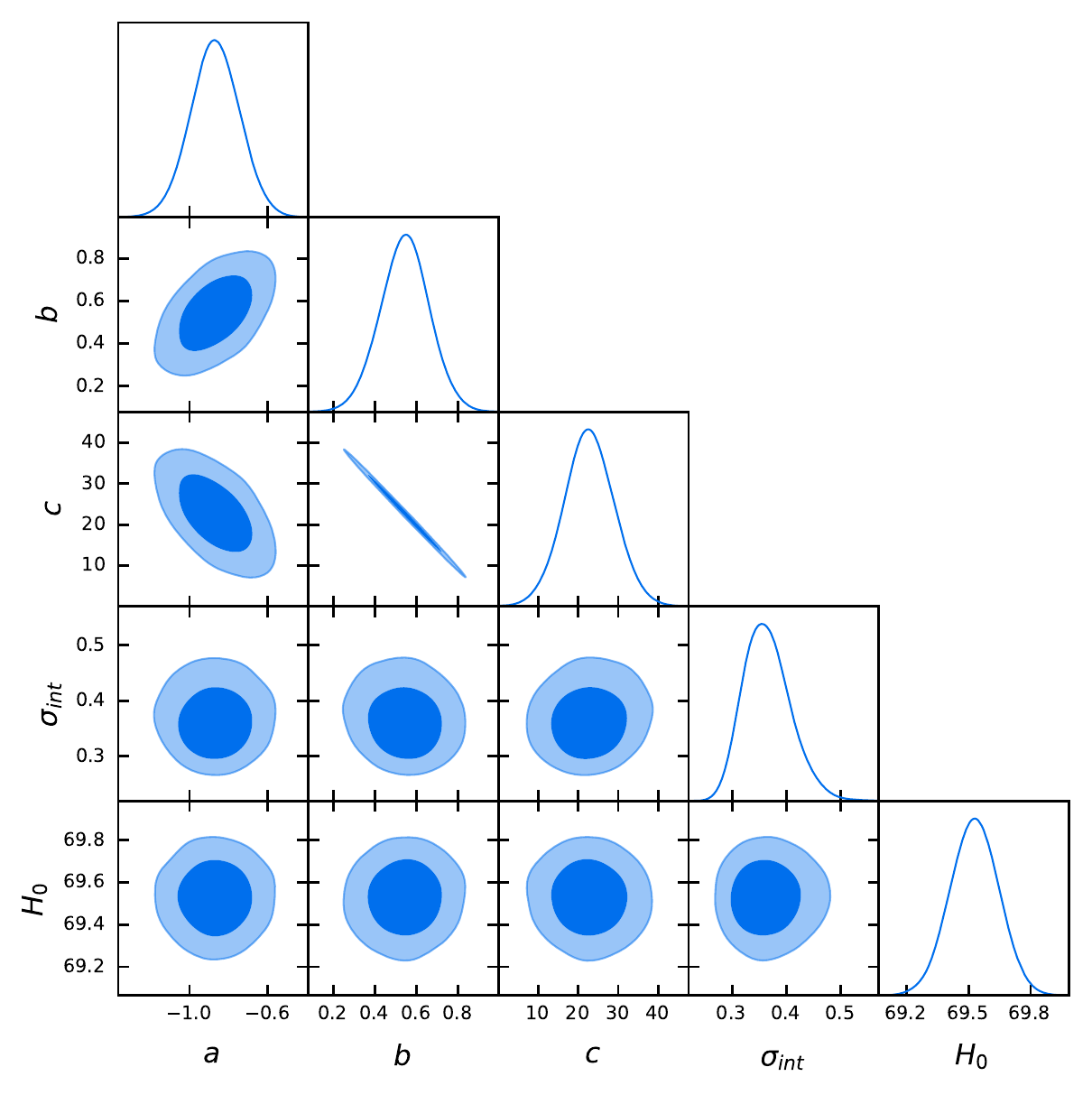}
\includegraphics[width=0.33\hsize,height=0.3\textwidth,angle=0,clip]{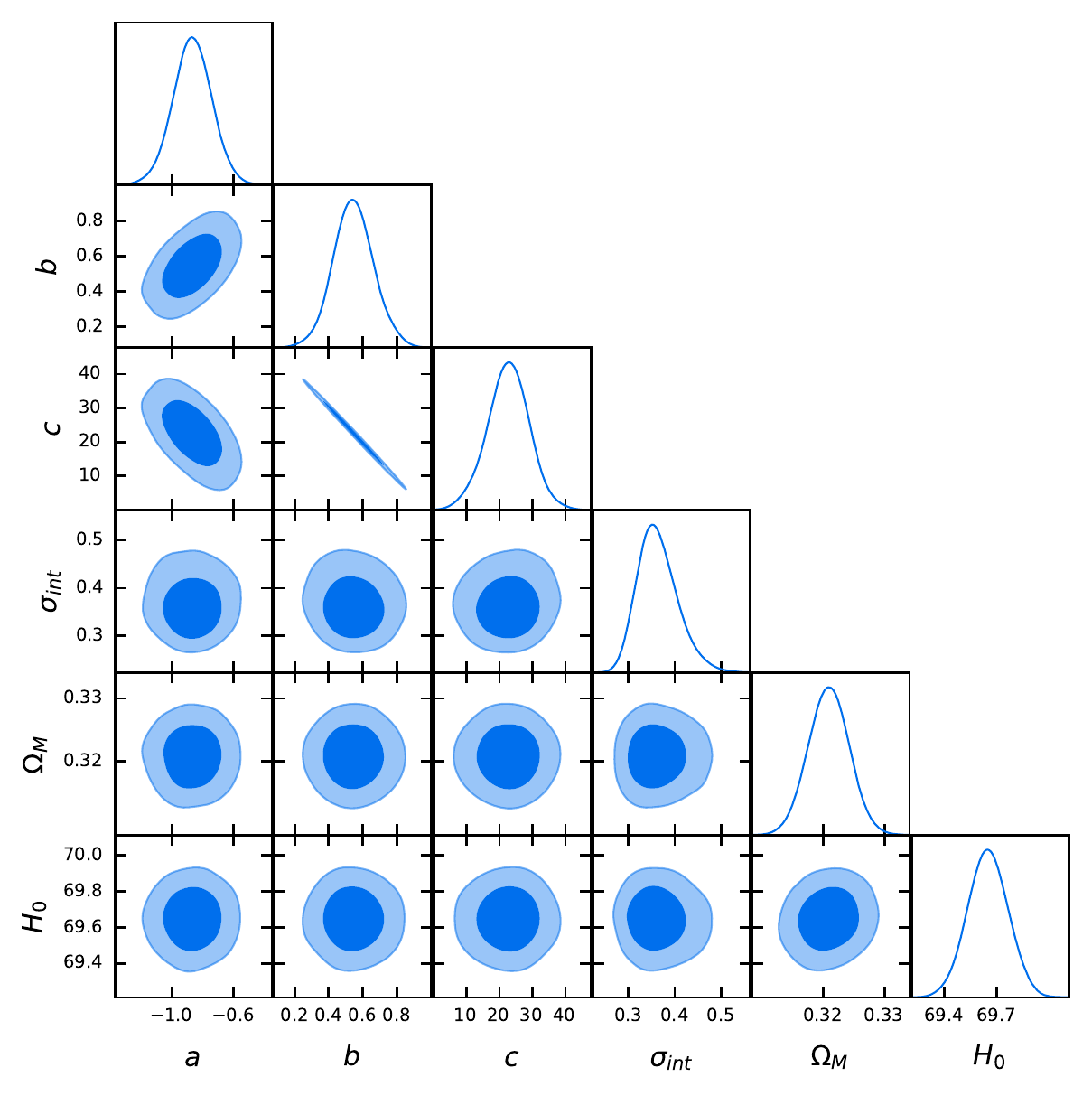}
\includegraphics[width=0.33\hsize,height=0.3\textwidth,angle=0,clip]{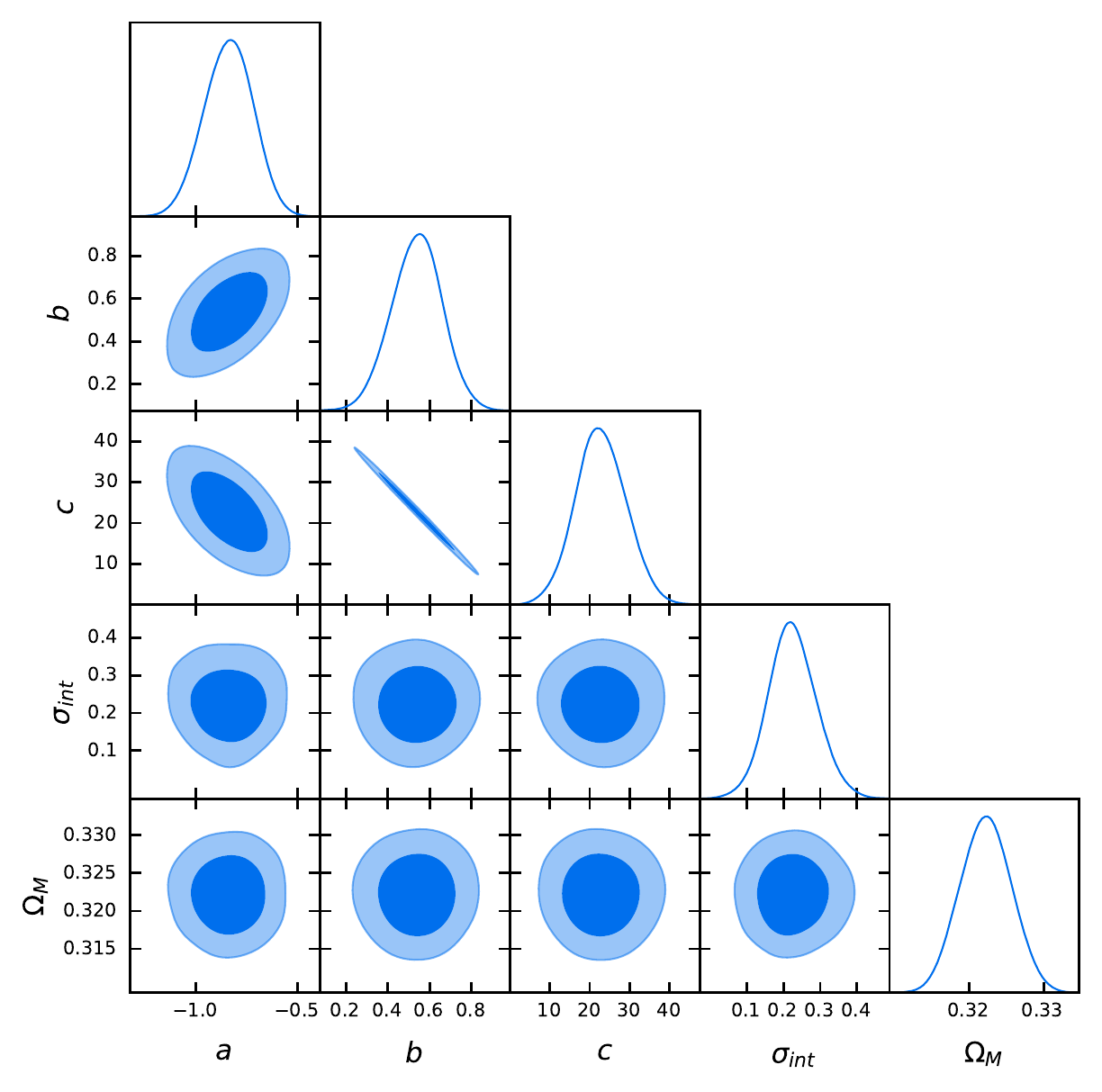}
\includegraphics[width=0.33\hsize,height=0.3\textwidth,angle=0,clip]{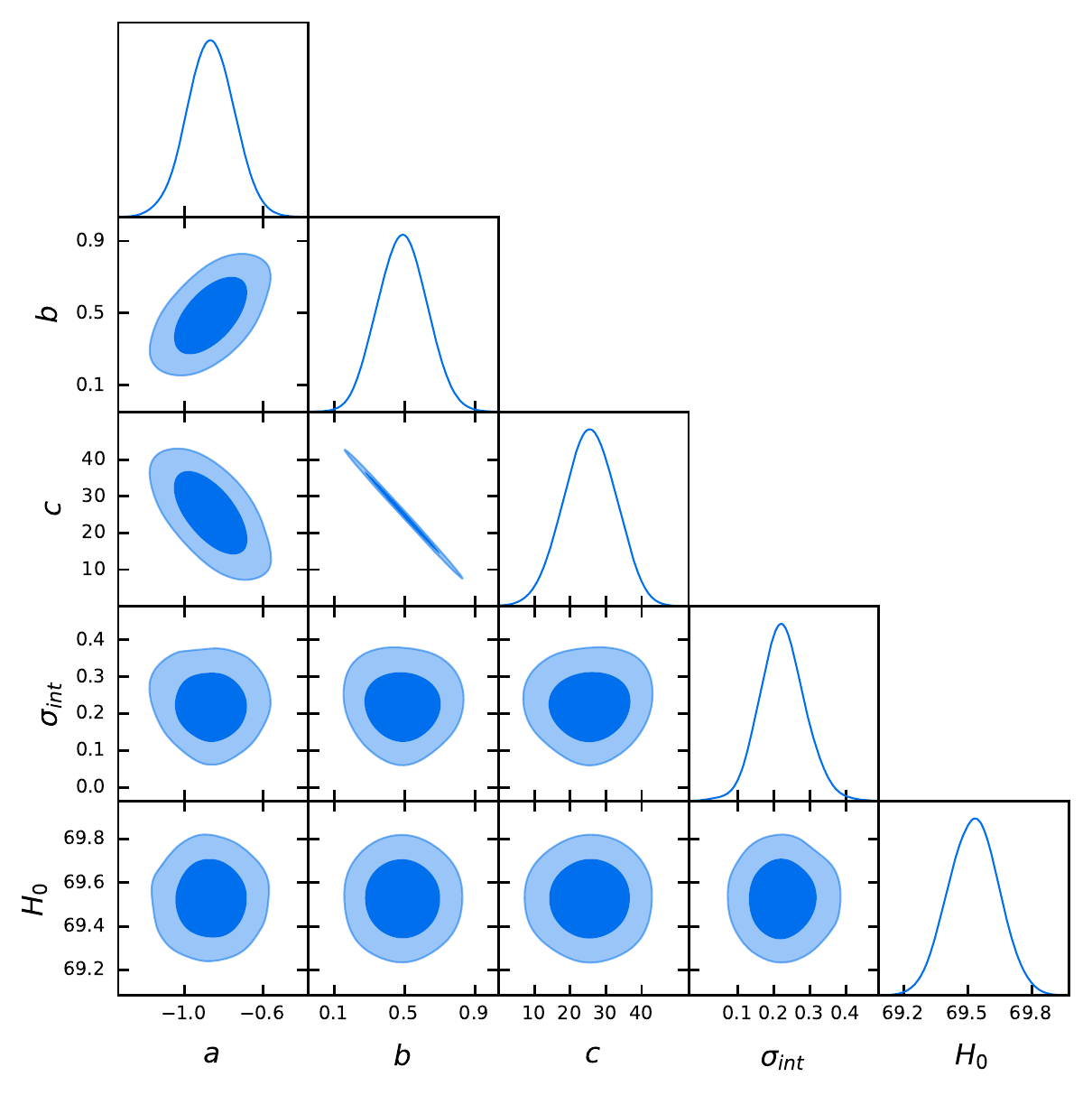}
\includegraphics[width=0.33\hsize,height=0.3\textwidth,angle=0,clip]{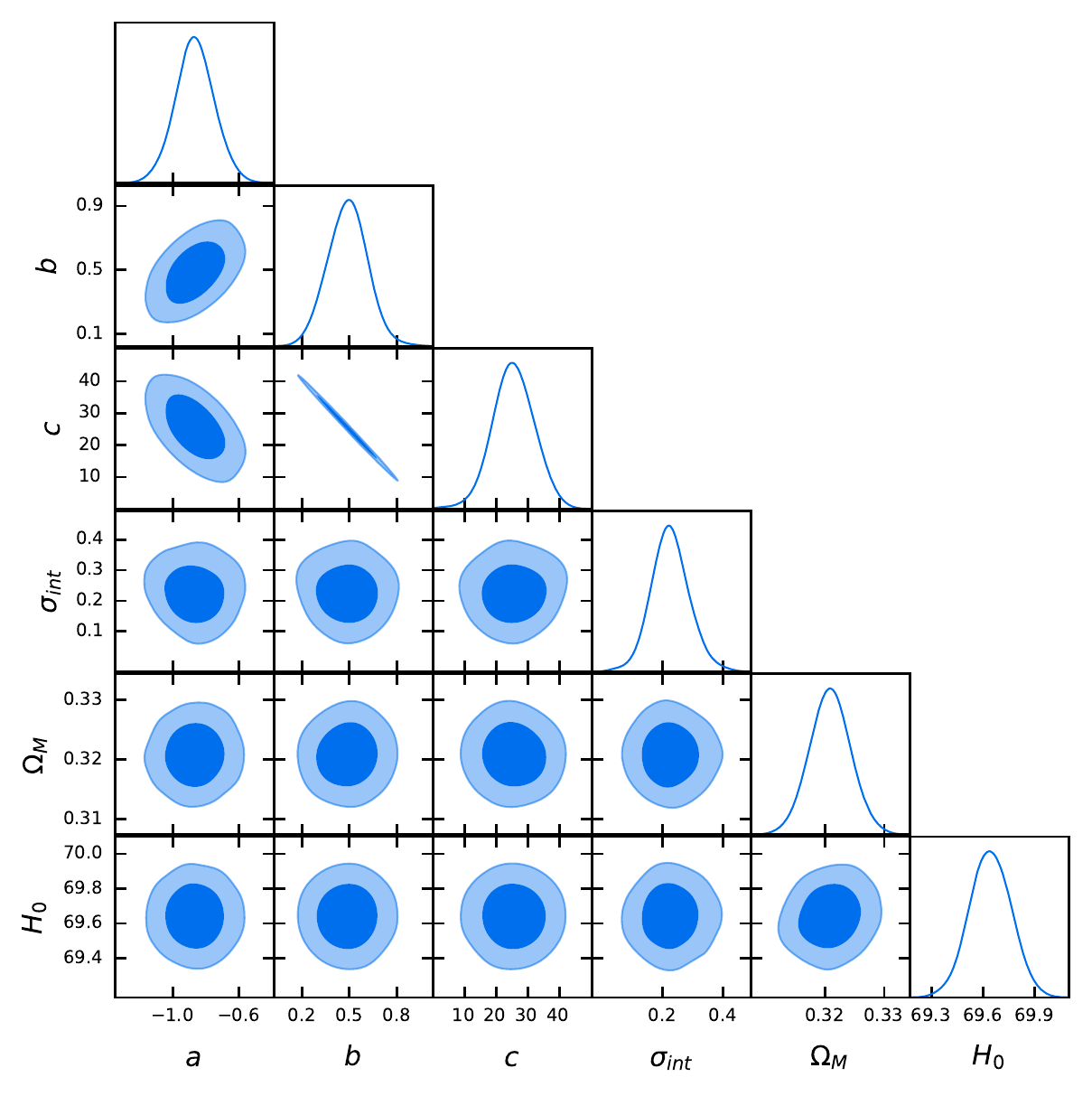}

\hspace{1 cm} 
\caption{Posterior contours computed for the full samples of SNe Ia, BAO and GRBs, considering the fundamental plane correlation for GRBs. On the upper panels the results without evolution are shown, varying $\Omega_{M}$ alone (upper left), $H_0$ alone (upper middle), and $\Omega_{M}$ and $H_0$ together (upper right). The same is shown in the bottom panels, where the evolutionary effects for GRBs are accounted for.}
\label{fundamentalplanecosmology}
\end{figure}

\section{Method for the derivation of cosmological parameters from GRB\lowercase{s}, SN\lowercase{e} Ia, and BAO\lowercase{s}} \label{Methodology Cosmology}

We here describe the methodology regarding the analysis of GRBs, SNe Ia, and BAOs to derive $H_0$ keeping $\Omega_M$ fixed at a fiducial value according to the $\Lambda$CDM model. Similarly, we infer $\Omega_M$ keeping $H_0$ fixed, and we derive both $\Omega_M$ and $H_0$ varying them contemporaneously. In particular, we present the computations performed to infer these cosmological parameters by using the fundamental plane correlation related to the PLAT sample both with and without the corrections due to the EP method. The GRB fundamental plane is used in combination with the SNe Ia data taken from the PS and the measurements obtained using the BAOs data, and we compute these cosmological parameters by considering the PS alone and SNe Ia+BAOs to make a comparison between the different sets.

We have performed a binned analysis of our data, dividing the PS into 5 subsets composed of 209 SNe Ia each (the only exception is the last bin 5, composed of 212 SNe Ia). The redshift range, as well as the mean redshift, for each bin are shown in Table 2. This binned analysis has the purpose to gather all GRBs into the last bin, given their higher redshift, to quantify their weight in the cosmological computations due to their inclusion.

Considering the flat $\Lambda$CDM standard model, fixing $\Omega_M=0.3$ and the equation of state $w=-1$, and neglecting the radiation contribution, the luminosity distance $d_L$ is:  
\begin{equation} \label{flatdistanceluminosity}
    d_L(z,\Omega_M, H_0)=(1+z)\frac{c}{H_{0}}\int_{0}^{z} \frac{dz'}{\sqrt{\Omega_{M}(1+z')^3+(1-\Omega_{M}})},
\end{equation}

\noindent where $c$ is the speed of light. We consider $H_0$ as a nuisance parameter contemporaneously with the other ones related to the fundamental plane correlation: in this way, varying $H_0$ together with these parameters allows us to not be biased by the so-called circularity problem. The same has been done when we compute $\Omega_M$ alone, and when we vary them both contemporaneously.
To evaluate the best value of $H_0$ describing the universe with a flat $\Lambda$CDM cosmology, we employ the distance moduli, $\mu_{obs,SNe}$, computed from the SNe Ia's observations.
We stress that, for the SNe Ia, equation (\ref{flatdistanceluminosity}) assumes the form:
\begin{equation}   d_L(z_{hel},z_{HD},\Omega_M, H_0)=(1+z_{hel})\frac{c}{H_{0}}\int_{0}^{z_{HD}} \frac{dz'}{\sqrt{\Omega_{M}(1+z')^3+(1-\Omega_{M}})},
\label{SNeDL}
\end{equation}
where $z_{hel}$ is the heliocentric redshift, and the integral is computed with the variable $z_{HD}$, namely the corrected redshift according to the cosmological rest frame \citep{Kenworthy}.



We first perform an analysis using GRBs+SNE Ia+BAOs for which we consider the fundamental GRB plane relation. We vary $\Omega_M$ or $H_0$ together with the fundamental plane parameters ($a$, $b$, $c$, and $\sigma_{int}$) using the D'Agostini method. The results are shown in figure $\ref{fundamentalplanecosmology}$.  We note that the intrinsic scatter computed is $\sigma_{int}= 0.34 \pm 0.04$, which is equal to the one inferred when the parameters of cosmology are fixed to the flat $\Lambda CDM$ model. For the case with evolution, we obtain $\sigma_{int}= 0.22 \pm 0.10$, which is again identical to the case when we fix the cosmological parameters.

We now explicitly derive the equation for $\mu_{obs}$ for GRBs related to the fundamental plane correlation that we are going to use for our cosmological computations together with SNe Ia and BAOs.
Without considering the evolutionary effects, after some algebraic manipulations we obtain the following equations: 
\begin{equation}
    \log_{10} L_X=\log_{10} F_{X}+\log_{10}(4\pi d_L^2 K_{X})=b \cdot \log_{10} (4\pi d_L^2 K_{peak})+b \cdot \log_{10} F_{peak}+a \cdot \log_{10} T^{*}_X+C,
\end{equation}

\noindent where $a,b,C$ are the coefficients of Equation (\ref{jetted}).
Then, we proceed by isolating the luminosity distance and, performing additional algebraic manipulation, we obtain:
\begin{equation}
\begin{split}
     \log_{10}(4\pi d_L^2)-b \cdot \log_{10}(4\pi d_L^2)=\\ b \cdot \log_{10} F_{peak}+a\cdot \log_{10} T^{*}_X+C-\log_{10} F_{X}-\log_{10}K_{X}+b \cdot \log_{10}K_{peak},
\end{split}
\end{equation}

Factorizing the distance luminosity, we obtain the following equation:

\begin{equation}
\begin{split}
    \log_{10}(d_L)=-\frac{\log_{10} F_{X}+\log_{10}K_{X}}{2 (1-b)}+\frac{b \cdot (\log_{10} F_{p}+\log_{10}K_{peak})}{2 (1-b)}\\ +\frac{(b-1)\log_{10}(4\pi)+C}{2 (1-b)}+ \frac{a \log_{10} T^{*}_X}{2 (1-b)}.
\end{split}
\end{equation}
Using equation (\ref{modulus}) and defining $a_1=-1/2(1-b)$; $b_1=b/2(1-b)$; $c_1=((b-1)\log_{10}(4\pi)+C)/2(1-b)$; $d_1=a/2 (1-b)$; $F_{peak,cor}= F_{p} \cdot K_{peak}$; and $F_{X,cor}= F_{X} \cdot K_{X}$,  we finally obtain the required equation:
\begin{equation}
\mu_{obs,GRBs}=5 (b_{1} \log_{10} F_{p,cor}+a_{1} \log_{10} F_{X,cor} + c_{1}+d_{1} \log_{10} T^{*}_X)+25,
\label{mu GRBs}
\end{equation}

\noindent where $K_{peak}$ and $K_X$ are the $K$-correction in the prompt and afterglow emission, respectively.
The same computations have been performed considering also the evolutionary functions introduced by the EP method, starting from equation (\ref{plane evolution}). 




The likelihood function for GRBs taking into account the $\mu_{obs, GRBs}$ can be written as follows:
\begin{equation}
\mathcal{L}_{GRB}=\sum_{i}\biggl(\log \biggl( \frac{1}{\sqrt{2\pi}\sigma_{\mu,i}} \biggr)- \frac{1}{2}\biggl( \frac{\mu_{th,GRB,i}-\mu_{obs,GRB,i}}{\sigma_{\mu,i}}\biggr)^{2} \biggr).
\label{Likelihood_GRB}
\end{equation}
\noindent where $\sigma_{\mu,i}$ has been computed by the error propagation of the observed quantities considering their logarithm in the following way:

\begin{equation}
\sigma_{\mu,i}=\sqrt{(5b_1 \cdot \log F_{p,err,cor})^2+(5a_1 \cdot \log F_{X,err,cor})^2+(5d_1 \cdot \log T_X^*)^2}.
    \label{sigmamu}
\end{equation}

\begin{table}
\centering

\label{BIN_division}
\begin{tabular}{|c|c|c|c|c|c|}
\hline
Bin & Mean $z$ & $z$ range & SNe & BAOs& GRBs \\\hline
1 & $0.038$ & $0.010 \le z \le 0.094$ &  209 & 0 & 0 \\\hline
2 & $0.156$ & $0.096 \le z \le 0.203$ &  209 & 2 & 0 \\\hline
3 & $0.248$ & $0.203 \le z \le 0.298$ & 209 & 0 & 0 \\\hline
4 & $0.375$ & $0.299 \le z \le 0.503$ &  209 & 3 & 0 \\\hline
5 & $1.071$ & $0.503 \le z \le 5$ &  212 & 11 & 50 \\\hline
\end{tabular}
\caption{The Table shows the number of bins, the average redshift of each bin, the number of SNe Ia, BAO, and GRBs given the division in the redshift ranges displayed in the second column.}
\end{table}

\begin{table}
\centering

\label{Full_sample_results}
\begin{tabular}{|c|c|c|c|c|c|}
\hline
Sample & $\Omega_{M}$ & $H_{0}$ &$\Omega_{M}$ and $H_{0}$   \\
& & $km \hspace{1ex} s^{-1} Mpc^{-1}$ & $km \hspace{1ex} s^{-1} Mpc^{-1}$     \\\hline
SNe Ia &   $0.299 \pm 0.009$ & $69.983 \pm 0.141$ & $0.298 \pm 0.020$ \\
& & & $70.014 \pm 0.313$  \\\hline
BAOs &   {\bf $0.326 \pm 0.004$} & \textbf{$68.172 \pm 0.227$} & \textbf{$0.286 \pm 0.015$} \\
& & & \textbf{$67.192 \pm 1.086$}  \\\hline
SNe Ia+BAOs &   \textbf{$0.322 \pm 0.003$} & $69.527 \pm 0.119$ & $0.321 \pm 0.003$ \\
& & & $69.644 \pm 0.121$  \\\hline
BAOs+GRBs NO EV &  $0.326 \pm 0.004$ & $68.172 \pm 0.224$ & $0.289 \pm 0.015$ \\
& & & $67.409 \pm 1.075$  \\\hline
BAO+GRBs  EV &   $0.326 \pm 0.004$ & $68.172 \pm 0.220$ & $0.286 \pm 0.015$ \\
& & & $67.219\pm 1.050$  \\\hline
BAOs+SNe Ia+GRBs NO EV &   $0.322 \pm 0.003$ & $69.527 \pm 0.119$ & $0.321 \pm 0.003$ \\
& & & $69.644 \pm 0.117$  \\\hline
BAOs+SNe Ia+GRBs EV &   $0.322 \pm 0.003$ & $69.527 \pm 0.113$ & $0.321 \pm 0.003$ \\
& & & $69.644 \pm 0.116$  \\\hline
\end{tabular}
\caption{ The Table shows the results of our cosmological computations for the full SN Ia, BAO, and GRB samples without the division in bins. With "EV" we indicate the correction for redshift evolution and selection biases regarding the GRB sample.}
\end{table}



 To compare the fundamental plane correlation with the observed data, we use equation (\ref{eq_luminosityplateau}) where this time we let the cosmological parameters vary together with the fundamental plane ones. 

Concerning the SNe Ia sample, the loglikelihood can be written as follows:

\begin{equation}
    \mathcal{L}_{SNe Ia}= \log  \frac{1}{\sqrt{2\pi |\mathcal{C}_{SNeIa}|}}-\frac{1}{2}((\mu_{th}-\mu_{obsSNe})^T\times \mathcal{C}_{SNe Ia}^{-1} \times (\mu_{th}-\mu_{obsSNe}),   \label{eq_chi2_SNe}
\end{equation}

\noindent where $\mathcal{C}_{SNe Ia}^{-1}$ is the inverted covariance matrix for the SNe Ia \citep{Scolnic}. Note that $\mathcal{C}_{SNe Ia}^{-1}$ is the inverse of every covariance submatrix, each of them related to a particular bin of the PS, introduced in section \ref{Methodology SNe}. Then, we have computed the cosmological parameters adopting only SNe Ia and BAOs data, to test if considering GRBs would lead to similar results even in the last bin, and to unveil if and how much we could increase our precision relatively to the results computed using only SNe Ia and BAO. Then, we have performed the cosmological computations both with and without corrections for selection biases to test how much our results change because of these corrections.

The most complete likelihood for our cosmological analysis is the following:

\begin{equation}
    \mathcal{L}_{SNe Ia+BAOs+GRBs}=\mathcal{L}_{SNe Ia}+\mathcal{L}_{BAO}+\mathcal{L}_{GRB}.
    \label{eq_likelihoodtotal}    
\end{equation}

\noindent We note that the likelihood in the form of equation (\ref{eq_likelihoodtotal}) is applied only to bin 5, where all of the GRBs are present, while for the other bins the likelihood takes the form $\mathcal{L}_{SNe Ia+BAOs}=\mathcal{L}_{SNe Ia}+\mathcal{L}_{BAO}$, with the exception of bins 1 and 3, where only the SNe Ia are present in these particular redshift ranges, which means we can only use $\mathcal{L}_{SNe Ia}$ for those. We report that the priors chosen for our Bayesian calculations regarding the cosmological parameters are uniform distributions: $0<\Omega_M<1$ and $60<H_0<80$. These priors have been used for all our computations.

\section{Results} \label{Results}
First we computed our cosmological parameters without the bin division, thus considering the full PS+ full PLAT+ full BAO samples. The results are shown in Table 3, while in figure \ref{wholesample} the posterior contours are shown regarding the BAOs+SNe Ia+GRBs cases, both with and without evolution, as well as for the BAO+GRBs case without evolution. Here we note that we have considered the following cases: 1) only SNe Ia 2) only BAOs; 3) SNe Ia+BAOs; 4) BAOs+GRBs without correcting for the evolutionary effects for GRBs; 5) BAOs+GRBs considering these effects; 6) BAOs+SNe Ia+GRBs without correction for evolutionary effects for GRBs and finally 7) BAOs+SNe Ia+GRBs with the correction for these effects. The results show that we obtain a better precision on the computed parameters when we combine more probes, in particular for the BAOs+SNe Ia+GRBs  sets both with and without the correction for the evolution in the case of GRBs. This is true for all the cosmological cases considered: varying only $\Omega_M$, varying only $H_0$, varying both $\Omega_M$ and $H_0$. We also note that the largest beneficial effect on the precision of the cosmological parameters comes from adding BAOs to the SNe Ia, but still a relevant effect on this precision can be noted once GRBs are taken into account, since the largest precision comes from the combination of the three probes together. The precision of the $\Omega_M$ parameter is the same when we consider SNe Ia+ BAO, or SNe+ BAO+GRBs both with and without accounting for evolutionary effects, and also in the cases in which we vary only $\Omega_M$ or when we vary both $\Omega_M$ and $H_0$ contemporaneously. Regarding $H_0$, instead, the most precise measurement comes from the combination of SNe Ia, BAO and GRBs both with and without evolutionary effects.

\begin{figure}
    \centering
    \includegraphics[width=0.55\hsize,height=0.5\textwidth,angle=0,clip]{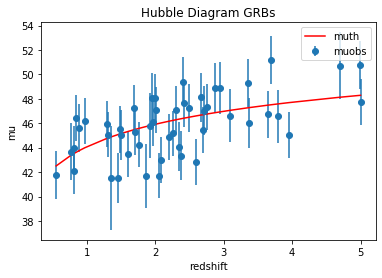}
    \caption{{\bf The Hubble Diagram computed considering only the platinum GRBs.}}
    \label{hubblediagram}
\end{figure}

\begin{figure}
\includegraphics[width=0.33\hsize,height=0.3\textwidth,angle=0,clip]{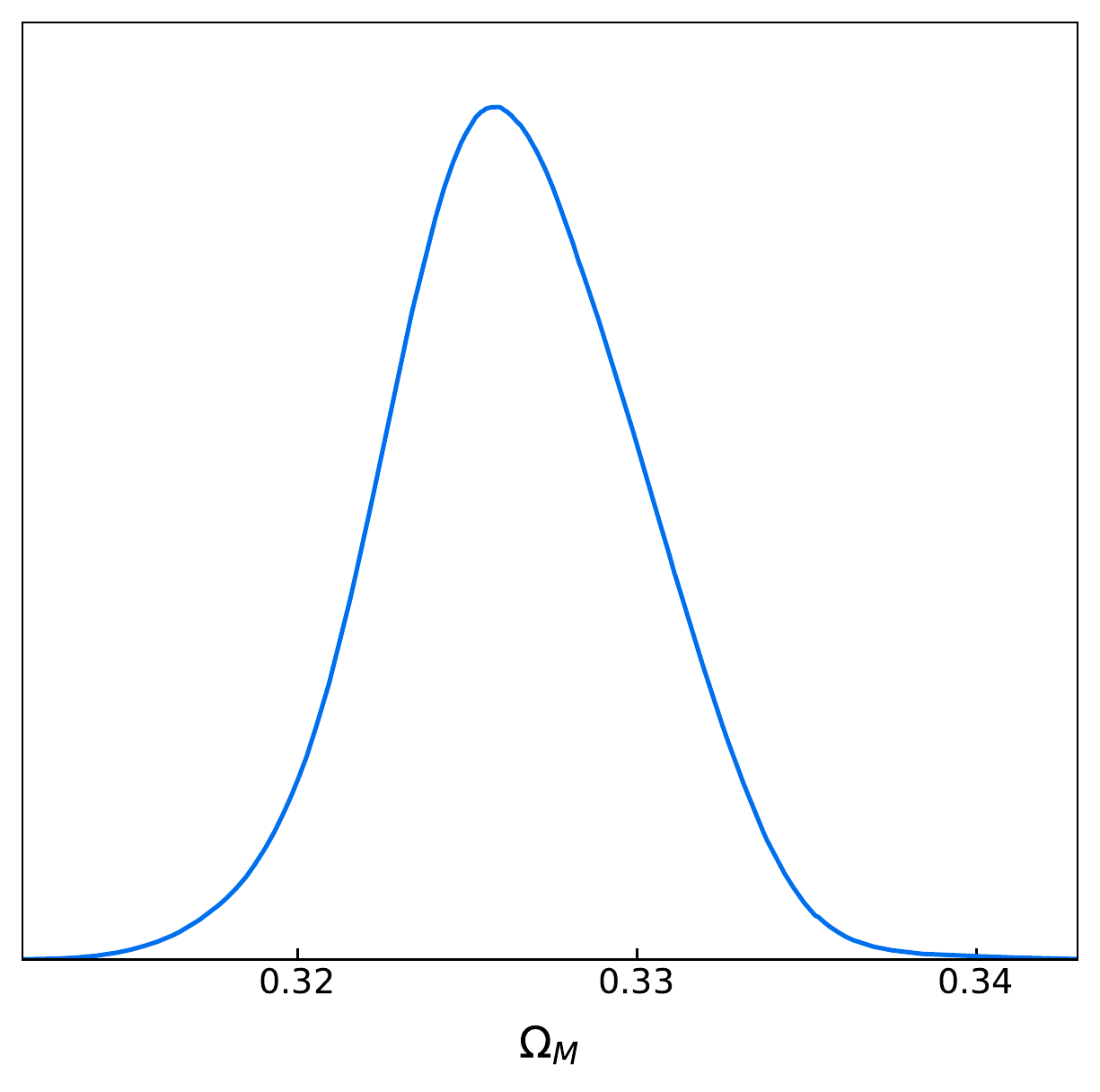}
\includegraphics[width=0.33\hsize,height=0.3\textwidth,angle=0,clip]{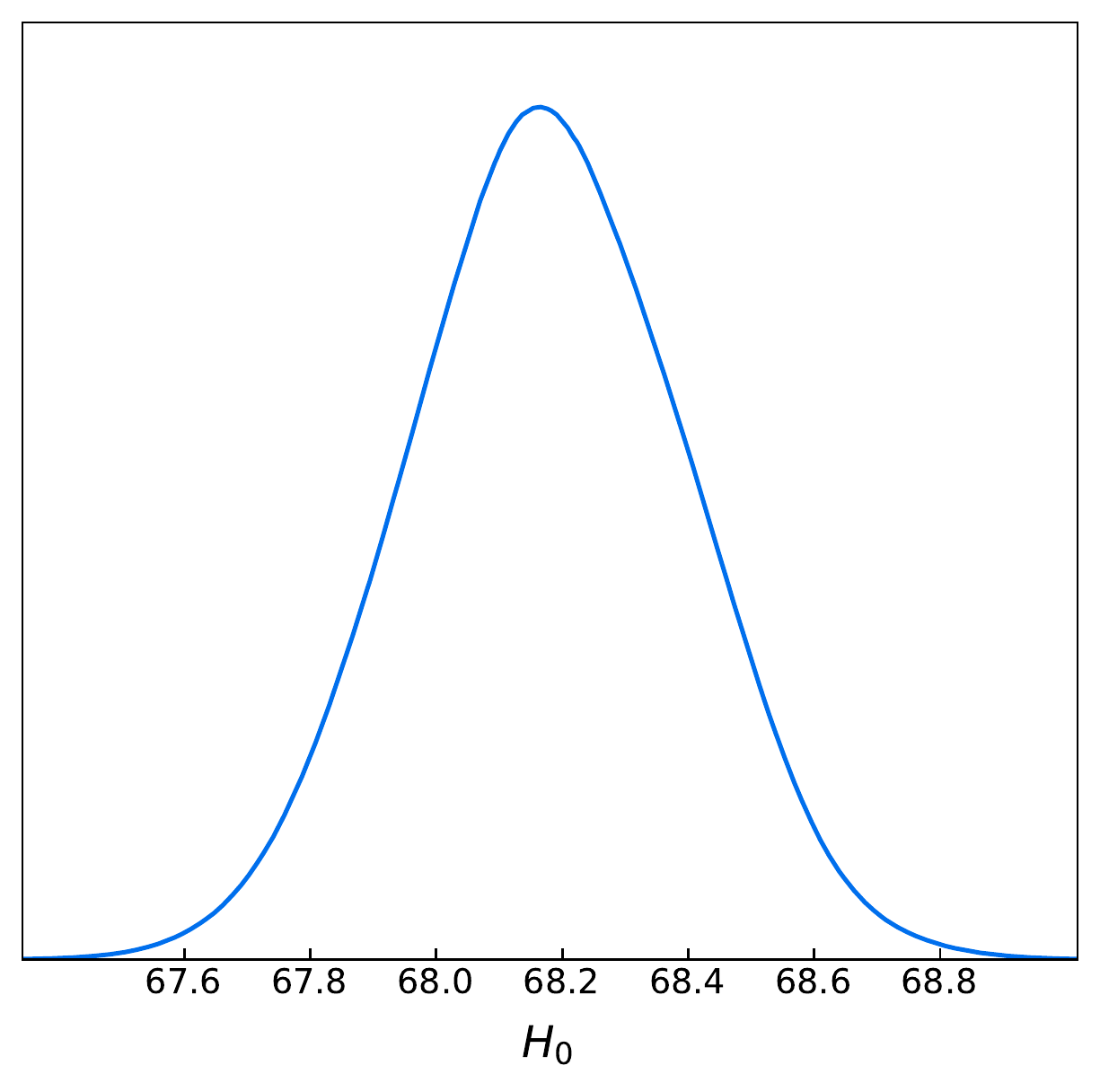}
\includegraphics[width=0.33\hsize,height=0.3\textwidth,angle=0,clip]{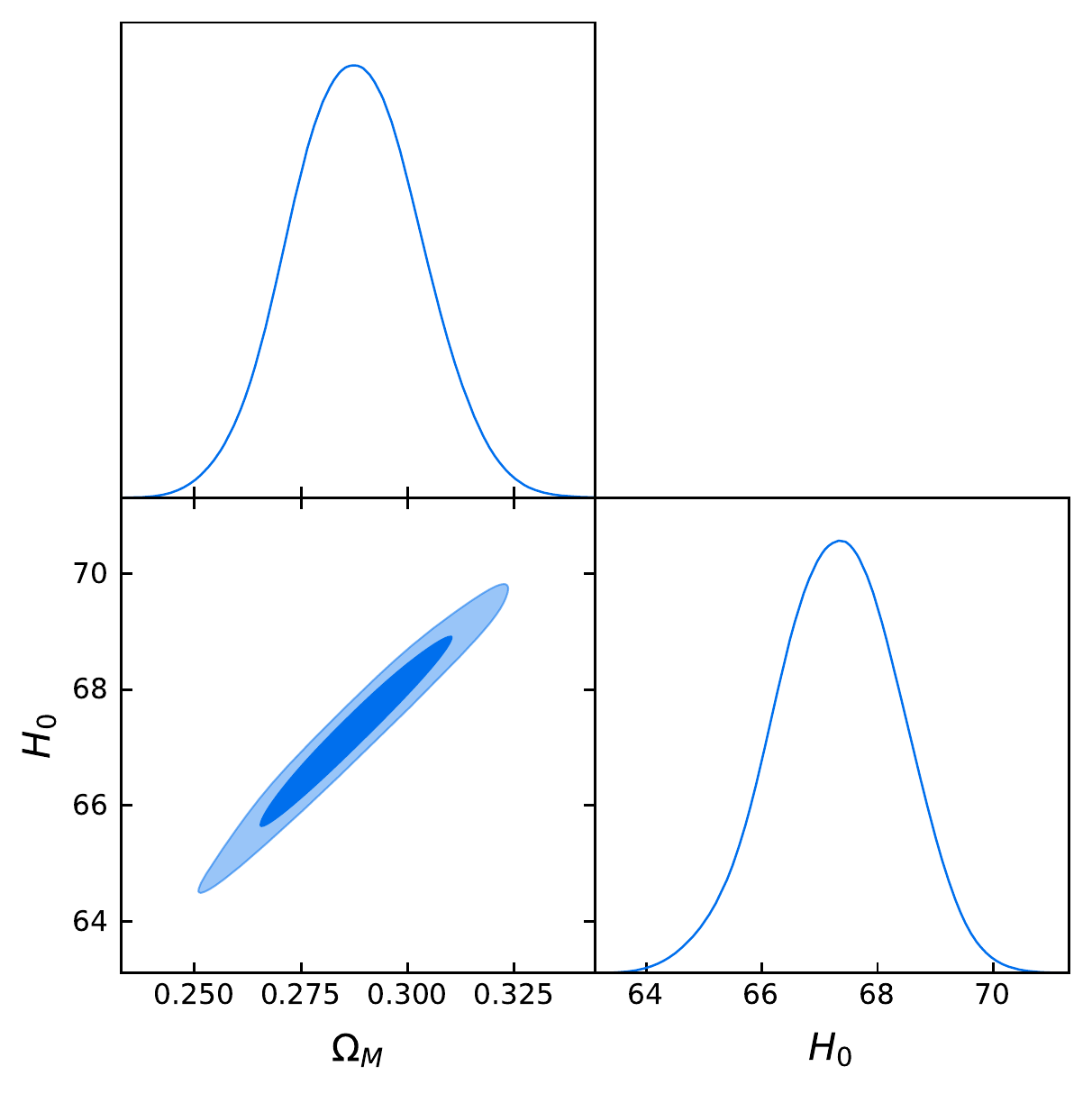}
\includegraphics[width=0.33\hsize,height=0.3\textwidth,angle=0,clip]{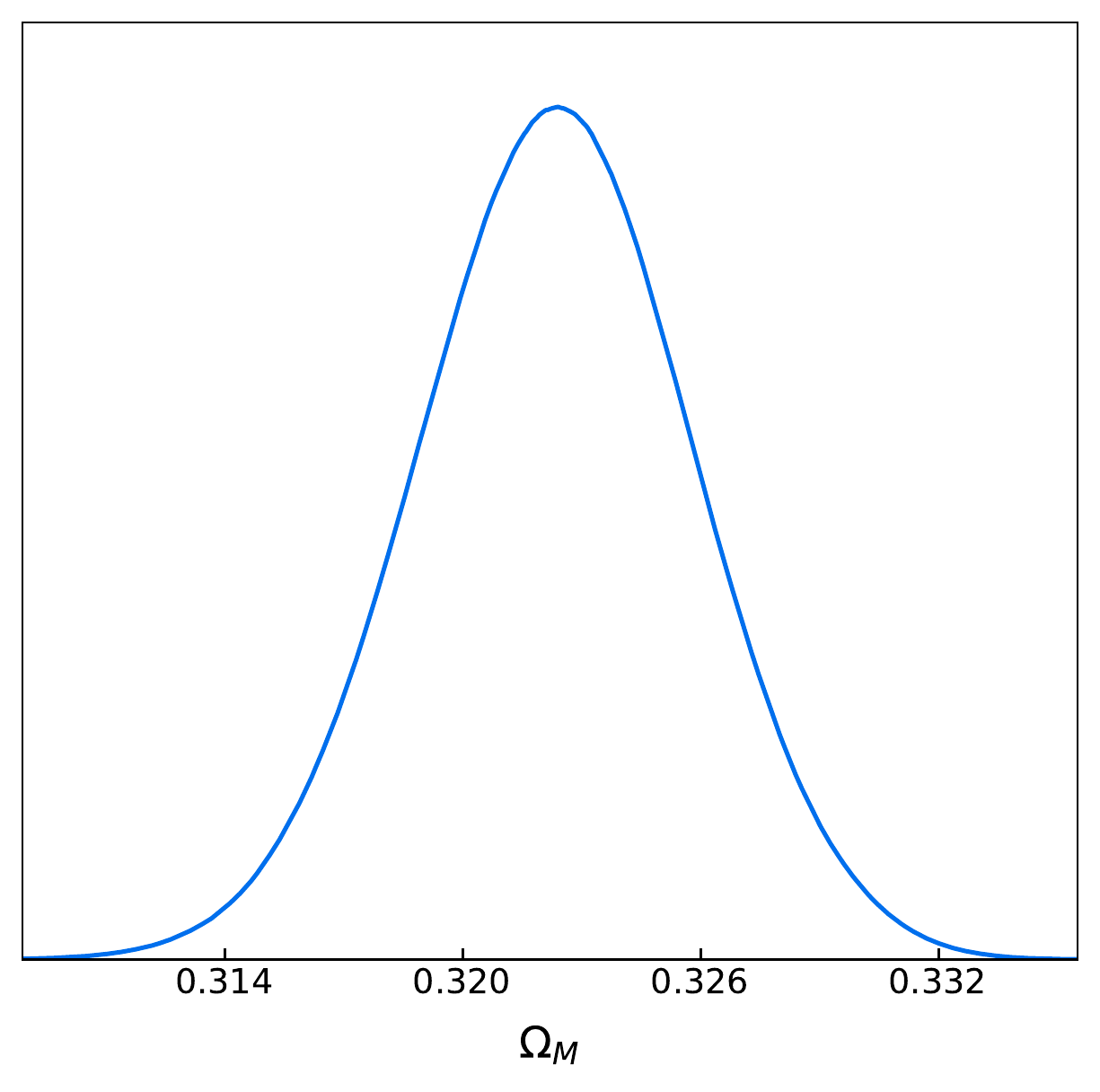}
\includegraphics[width=0.33\hsize,height=0.3\textwidth,angle=0,clip]{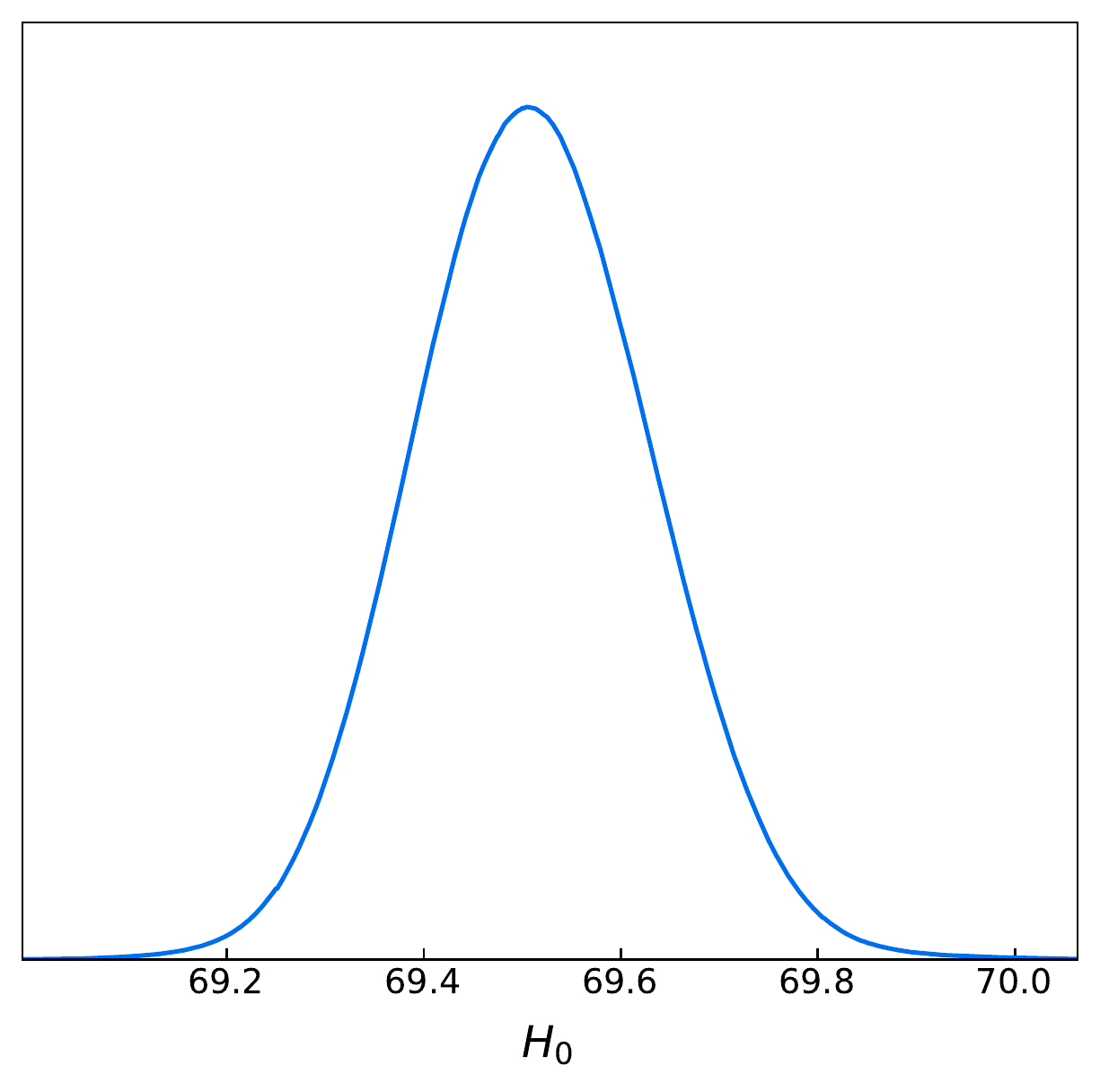}
\includegraphics[width=0.33\hsize,height=0.3\textwidth,angle=0,clip]{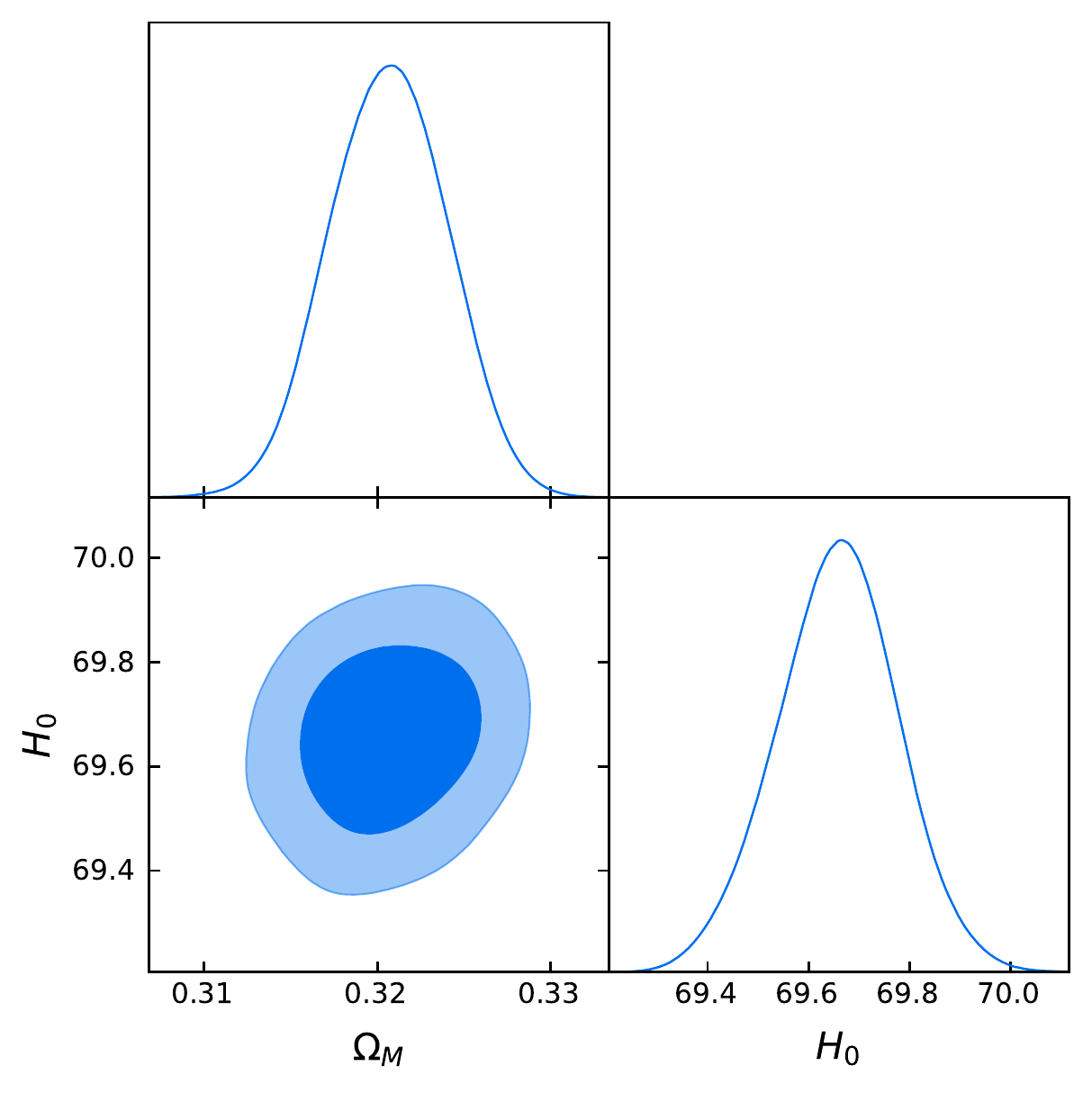}
\includegraphics[width=0.33\hsize,height=0.3\textwidth,angle=0,clip]{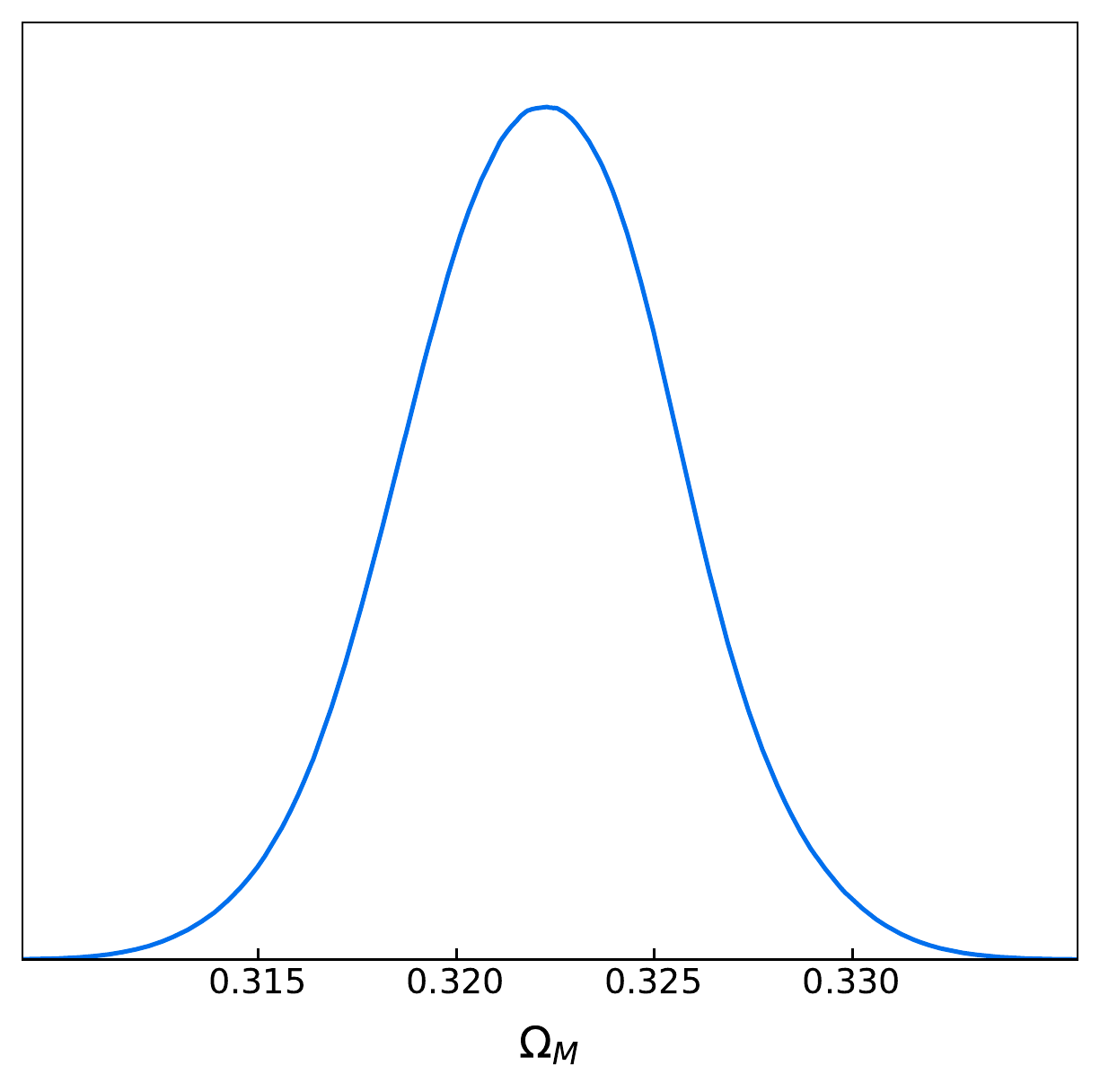}
\includegraphics[width=0.33\hsize,height=0.3\textwidth,angle=0,clip]{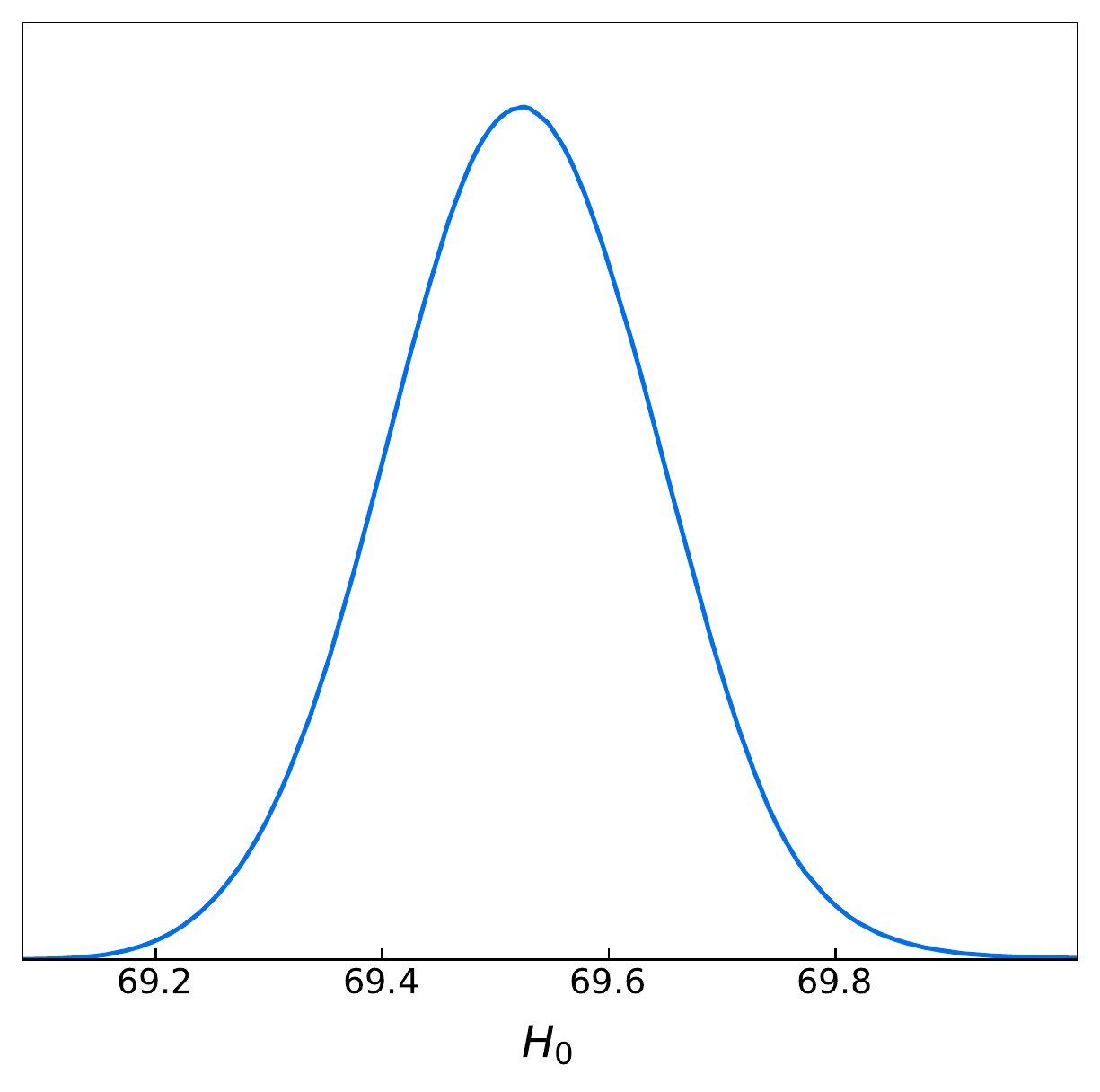}
\includegraphics[width=0.33\hsize,height=0.3\textwidth,angle=0,clip]{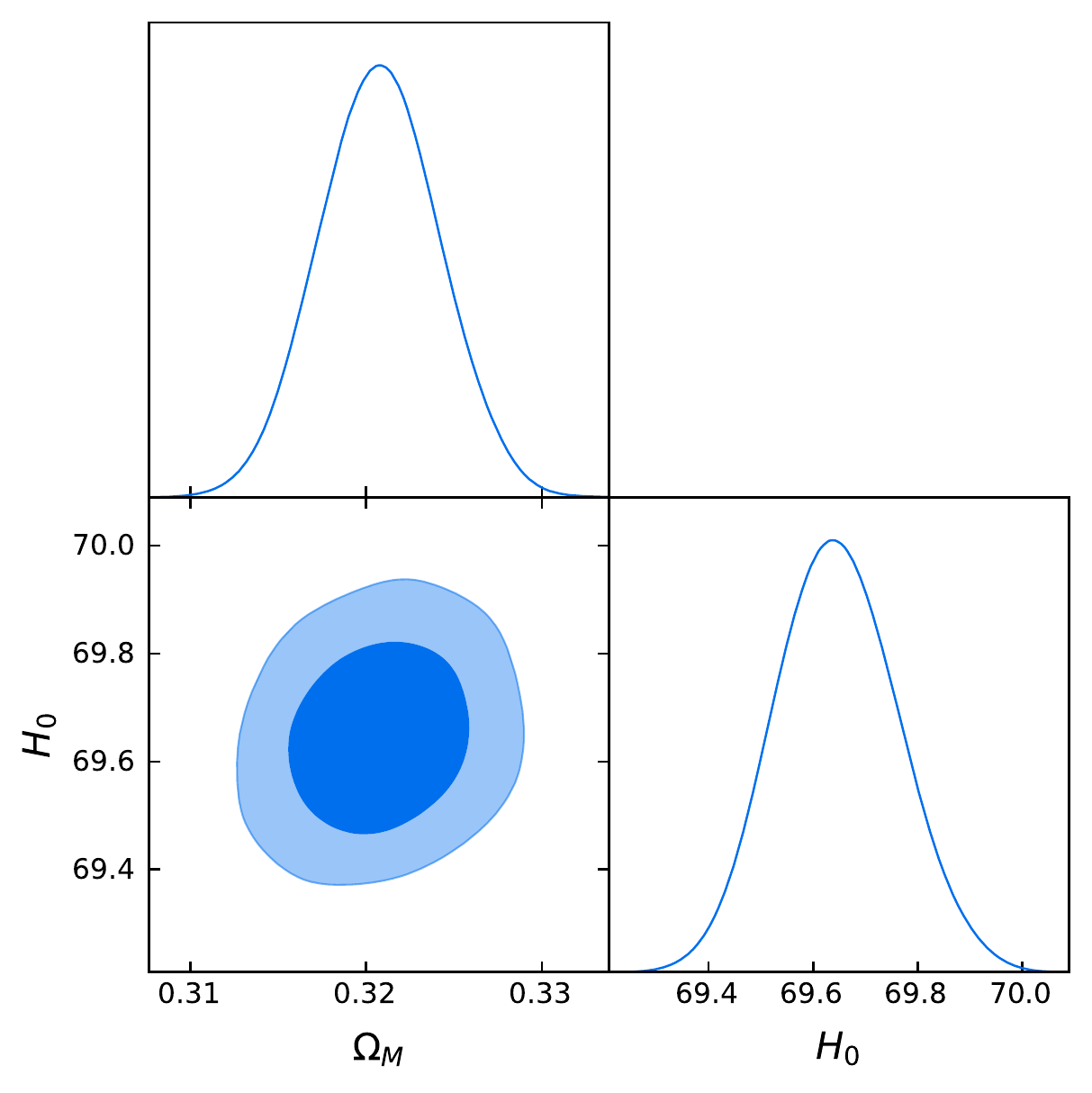}

\caption{ Posterior contours computed for the full samples of SNe Ia, BAOs, and GRBs. On the upper panels the results for BAOs+GRBs without evolution are shown varying $\Omega_{M}$ alone (upper left), $H_0$ alone (upper middle), and $\Omega_{M}$ and $H_0$ together (upper right). On the central panels, instead, the results for SNe Ia+BAOs+GRBs without evolution are shown, in the same order as before. The same is shown in the bottom panels, where the evolutionary effects for GRBs are taken into account.}
\label{wholesample}
\end{figure}

 We also show in figure \ref{hubblediagram} the Hubble diagram obtained by us considering only the GRBs belonging to the PLAT sample, computed for the case where we vary $\Omega_M$ without evolution. We note that the majority of the distance moduli values obtained using equation (16) with the best fit parameters derived by us are consistent with the theoretical distance modulus computed using the luminosity distance formula reported in equation (11). 
We also note that a few data points are not consistent with the best-fit curve in 1 $\sigma$. This could be due to large dispersion effects that may be related to the errors on the measured quantities in our fit, but still, given the majority of the data points are well-fitted by a given fiducial cosmological model, this shows a good hint for the reliability of GRBs as cosmological tools.

 Next, we apply the bin division for our samples. The numbers of all the objects in each bin are shown in Table 2. We note that, given the redshift range, we find SNe Ia in all bins, while all GRBs are gathered inside bin 5, as anticipated in previous sections. BAOs are found in  bins 2, 4, 5; we also note that the majority of the BAO measurements belong to bin 5. This implies that we have only SNe Ia for bins 1 and 3, SNe Ia and BAOs for bins 2, 4, and SNe Ia+BAOs+GRBs for bin 5.

\begin{figure}
\includegraphics[width=0.5\hsize,height=0.4\textwidth,angle=0,clip]{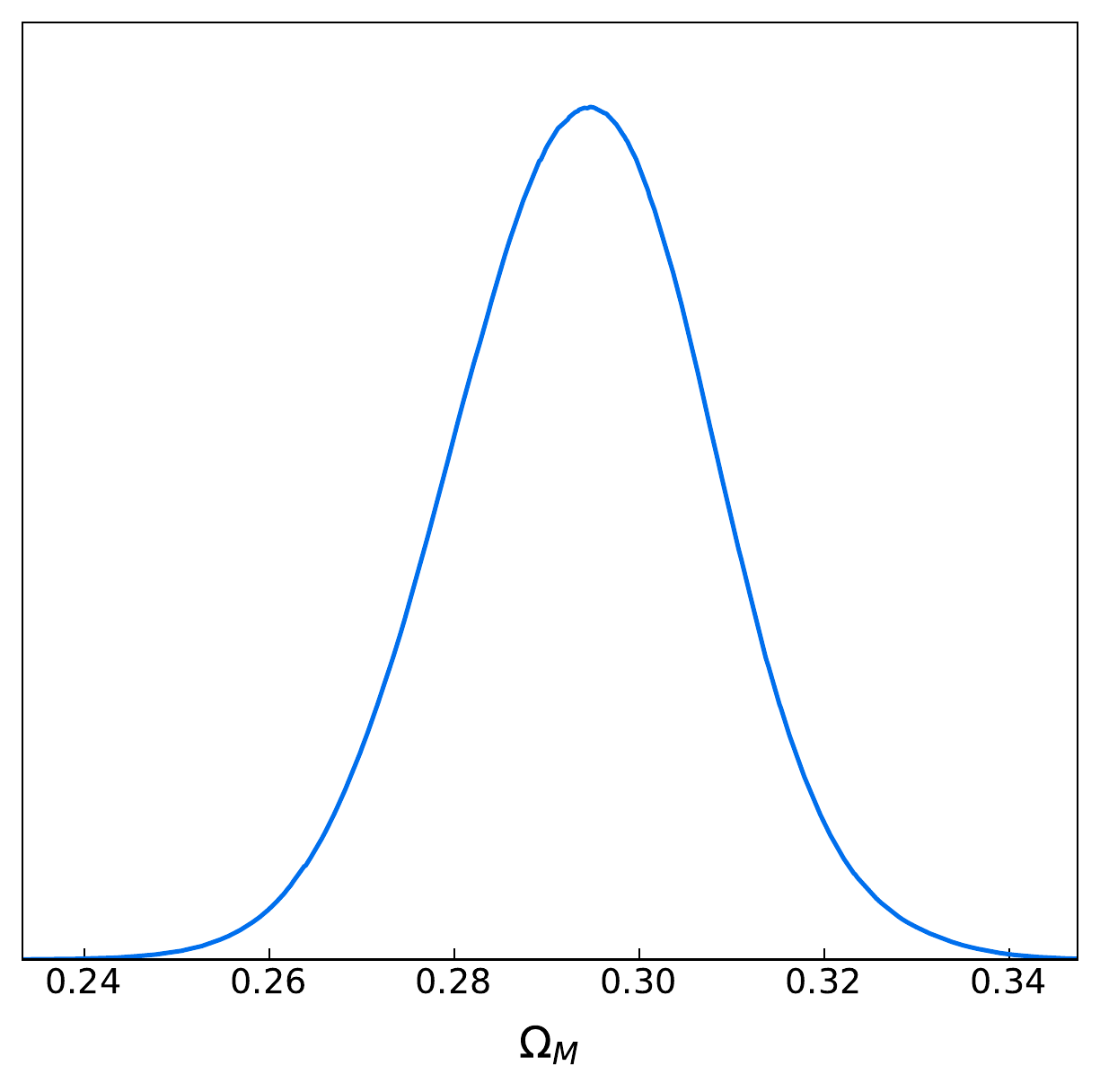}
\includegraphics[width=0.5\hsize,height=0.4\textwidth,angle=0,clip]{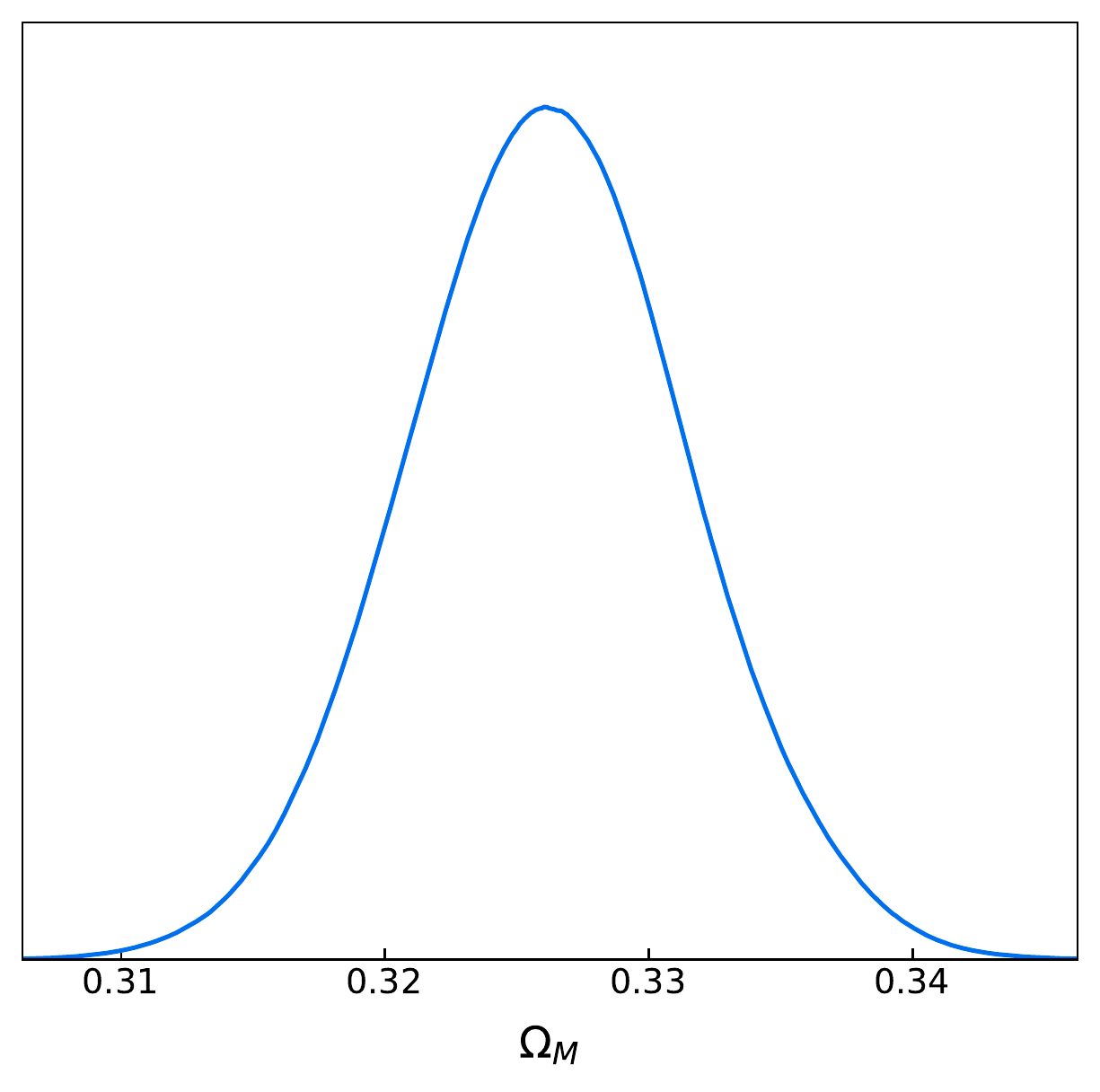}
\includegraphics[width=0.5\hsize,height=0.4\textwidth,angle=0,clip]{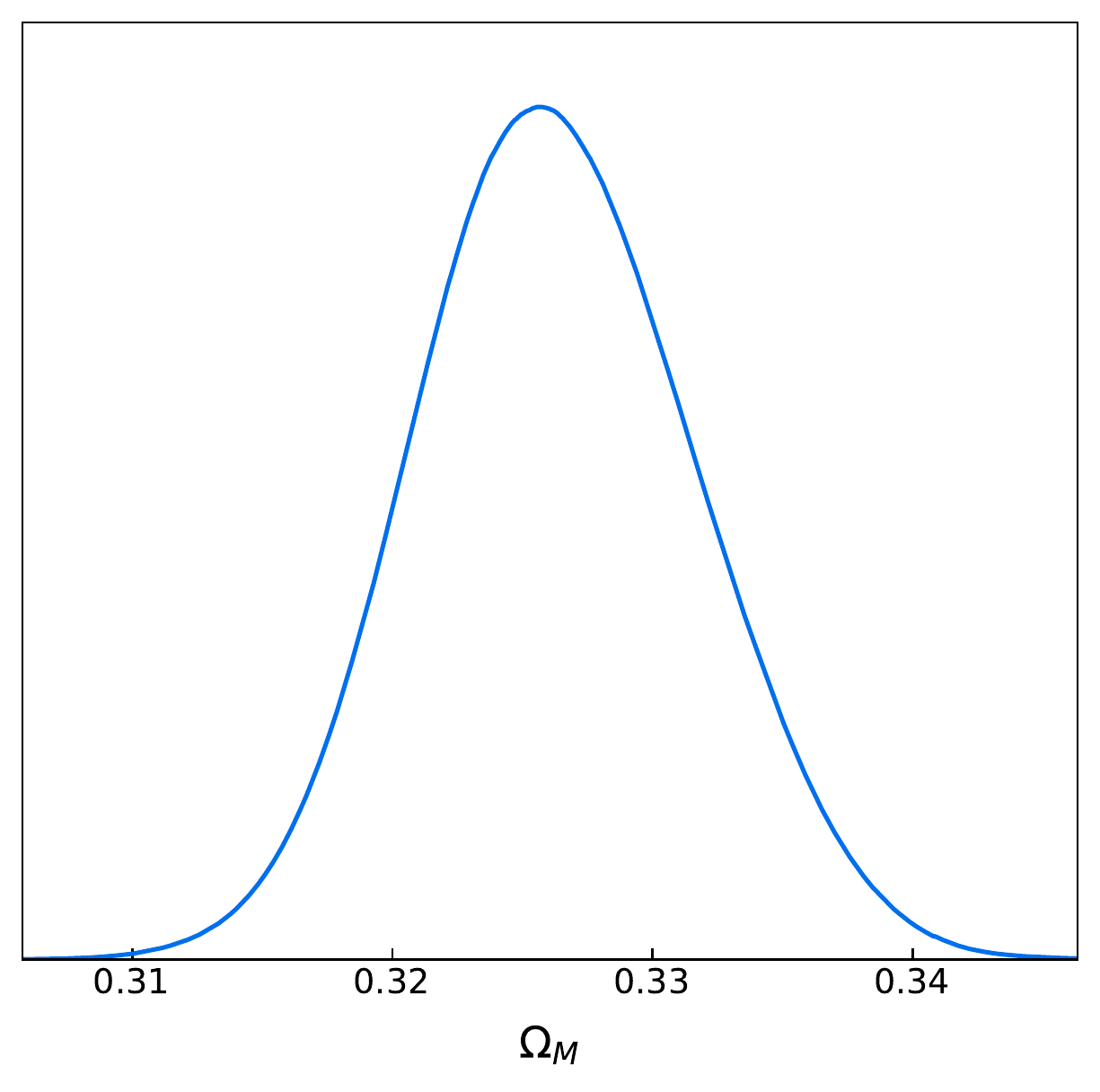}
\includegraphics[width=0.5\hsize,height=0.4\textwidth,angle=0,clip]{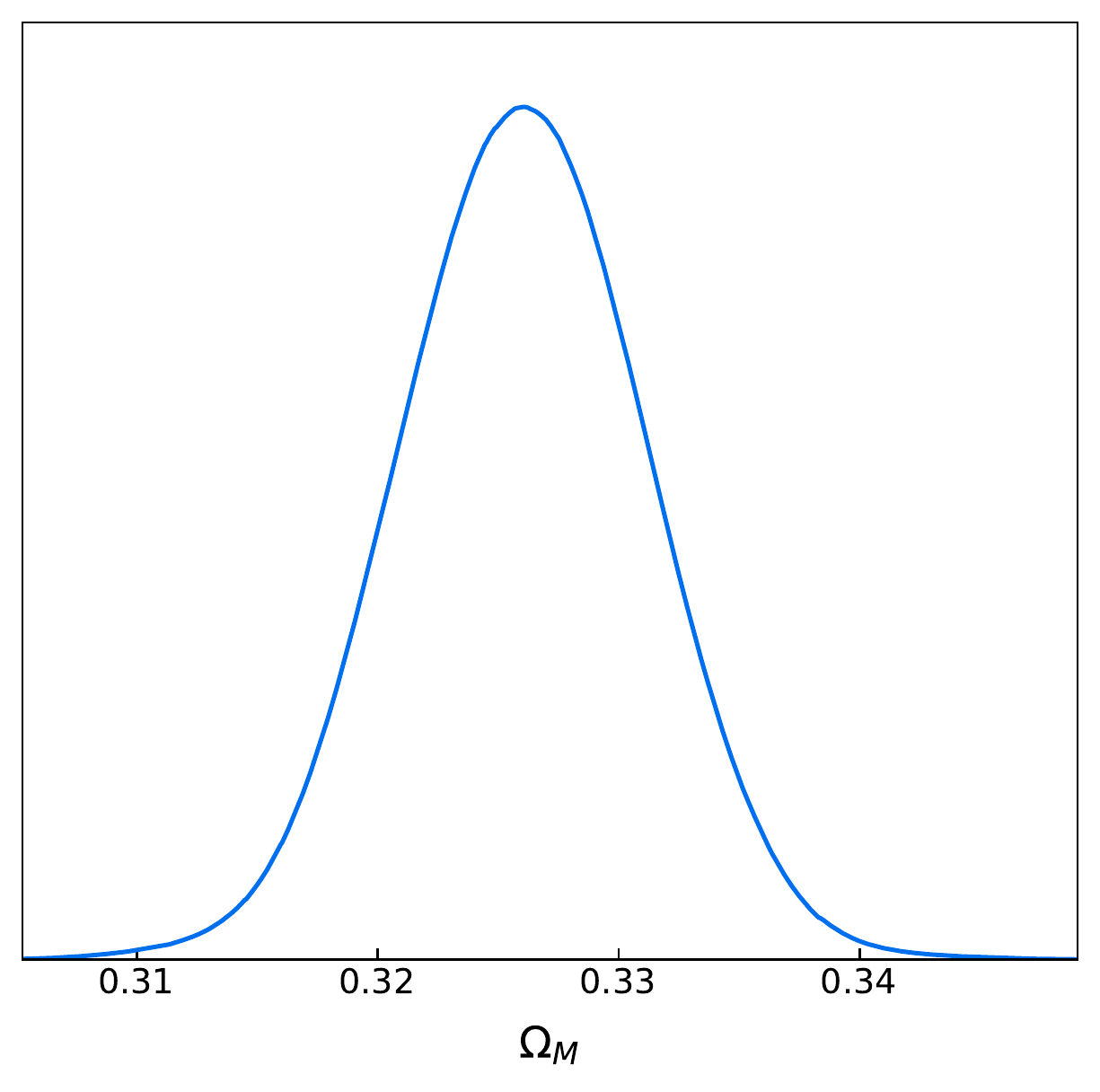}

\caption{\textbf{Results of the cosmological computations of bin 5, varying only $\Omega_{M}$, for SNe Ia (top left panel) SNe Ia+BAOs (top right panel), and GRBs+SNe IA+BAOs,  both without (bottom left panel) and with (bottom right panel) considering the evolutionary effects.}}
\label{OMBIN5}
\end{figure}

\begin{figure}
\includegraphics[width=0.5\hsize,height=0.4\textwidth,angle=0,clip]{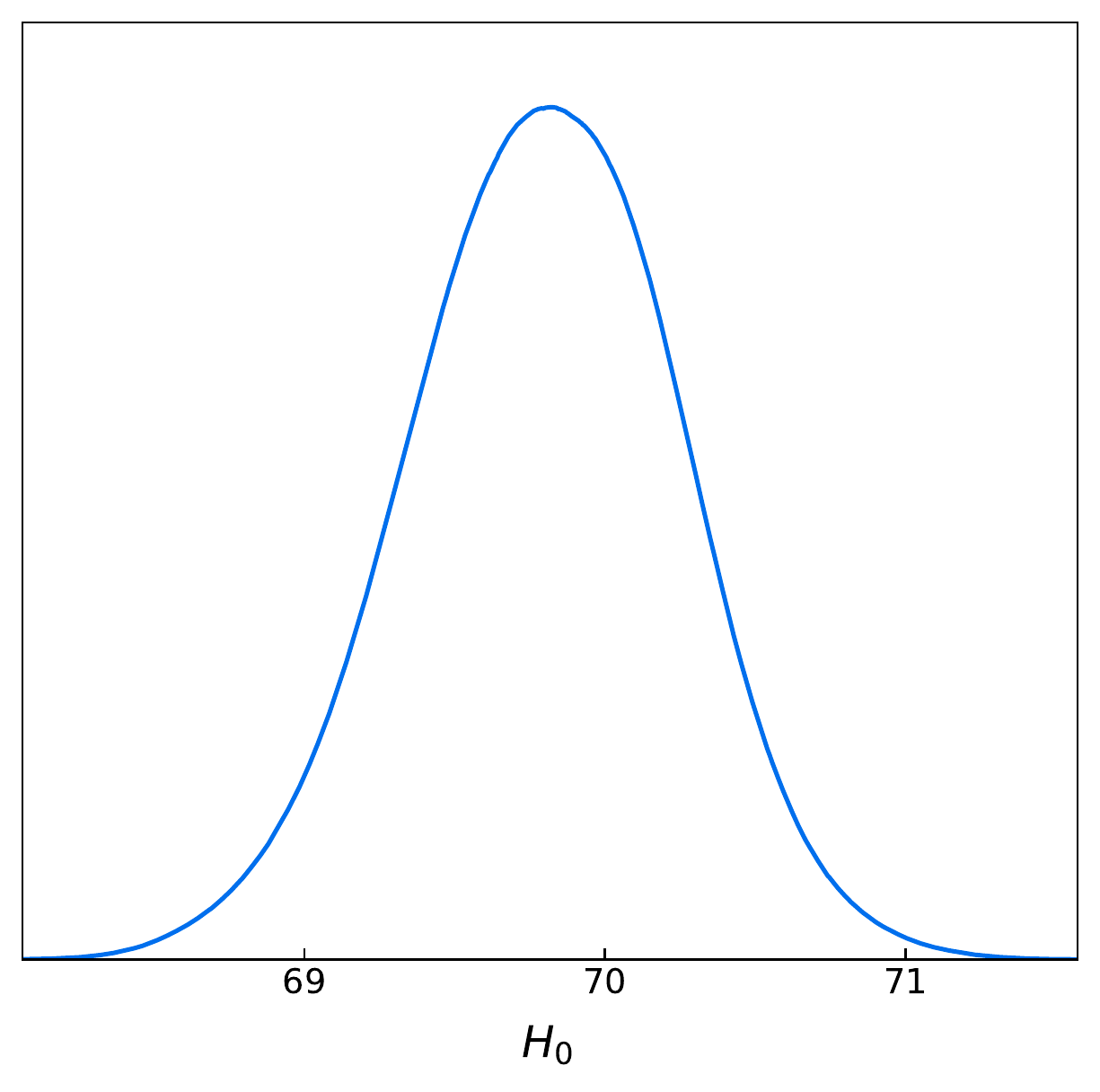}
\includegraphics[width=0.5\hsize,height=0.4\textwidth,angle=0,clip]{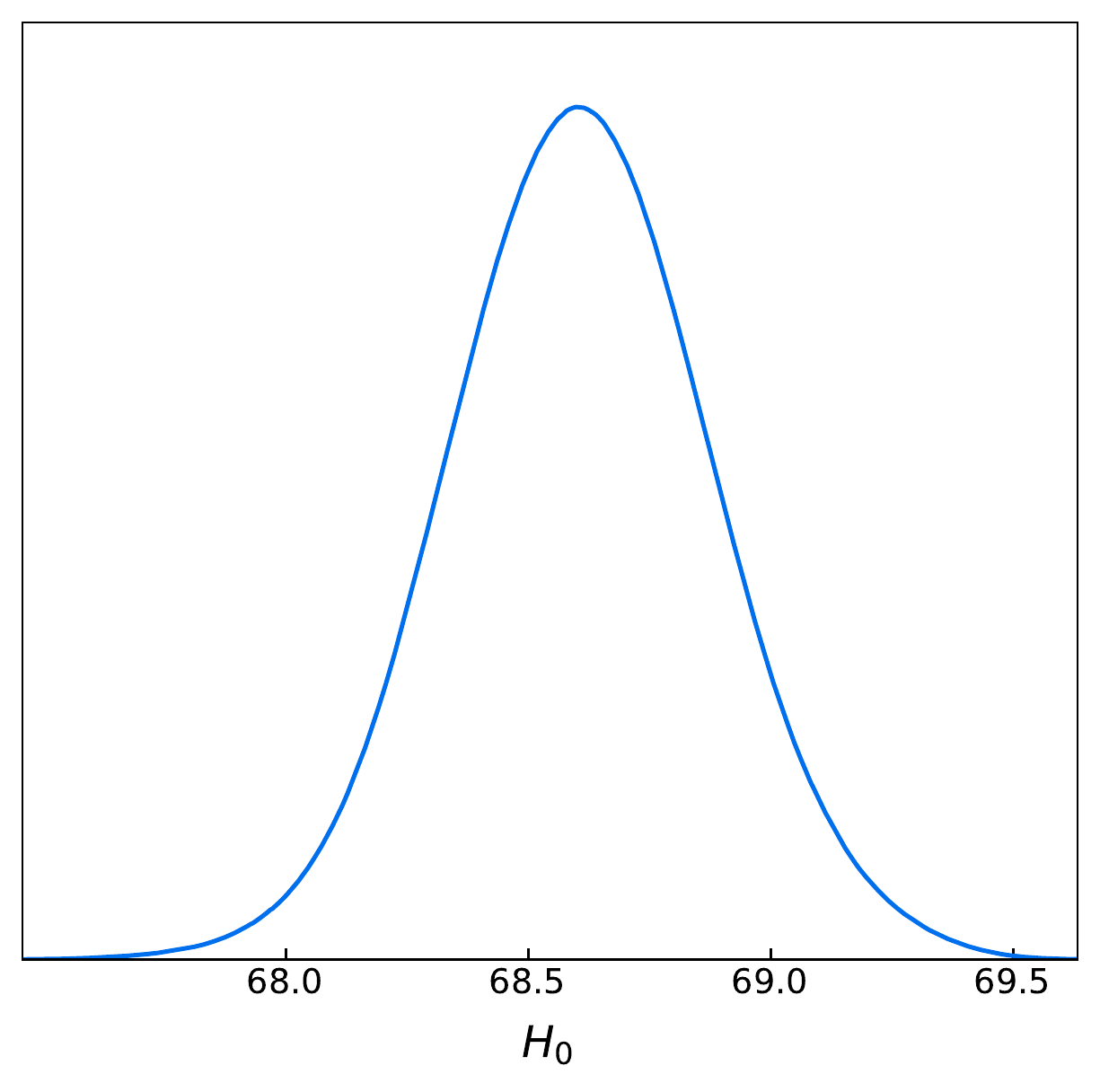}
\includegraphics[width=0.5\hsize,height=0.4\textwidth,angle=0,clip]{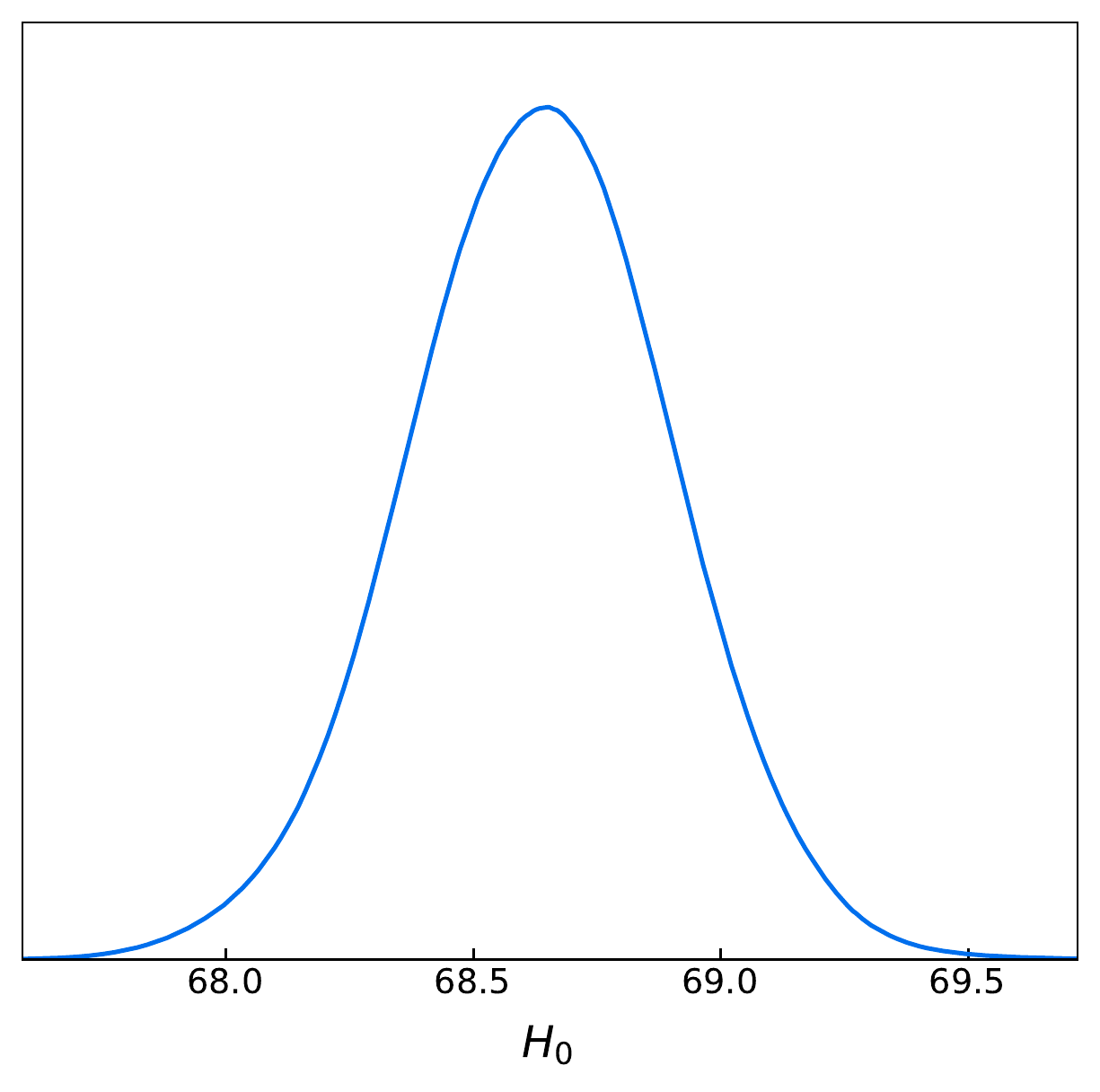}
\includegraphics[width=0.5\hsize,height=0.4\textwidth,angle=0,clip]{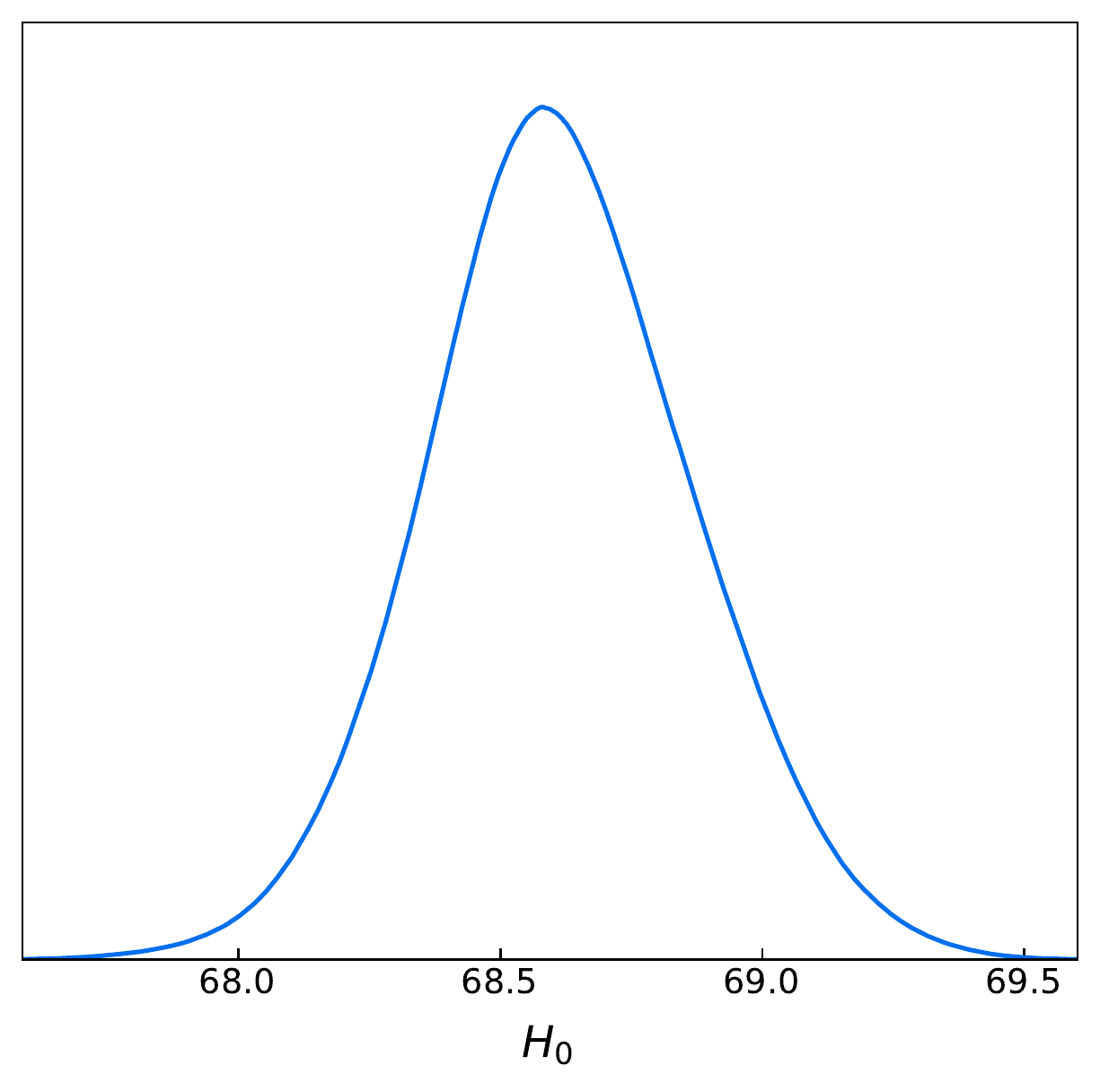}

\caption{\textbf{Results of the cosmological computations of bin 5, varying only $H_0$, for SNe Ia (top left panel) SNe Ia+BAOs (top right panel), and GRBs+SNe IA+BAOs,  both without (bottom left panel) and with (bottom right panel) considering the evolutionary effects.}}
\label{H0BIN5}
\end{figure}

\begin{figure}
\includegraphics[width=0.5\hsize,height=0.4\textwidth,angle=0,clip]{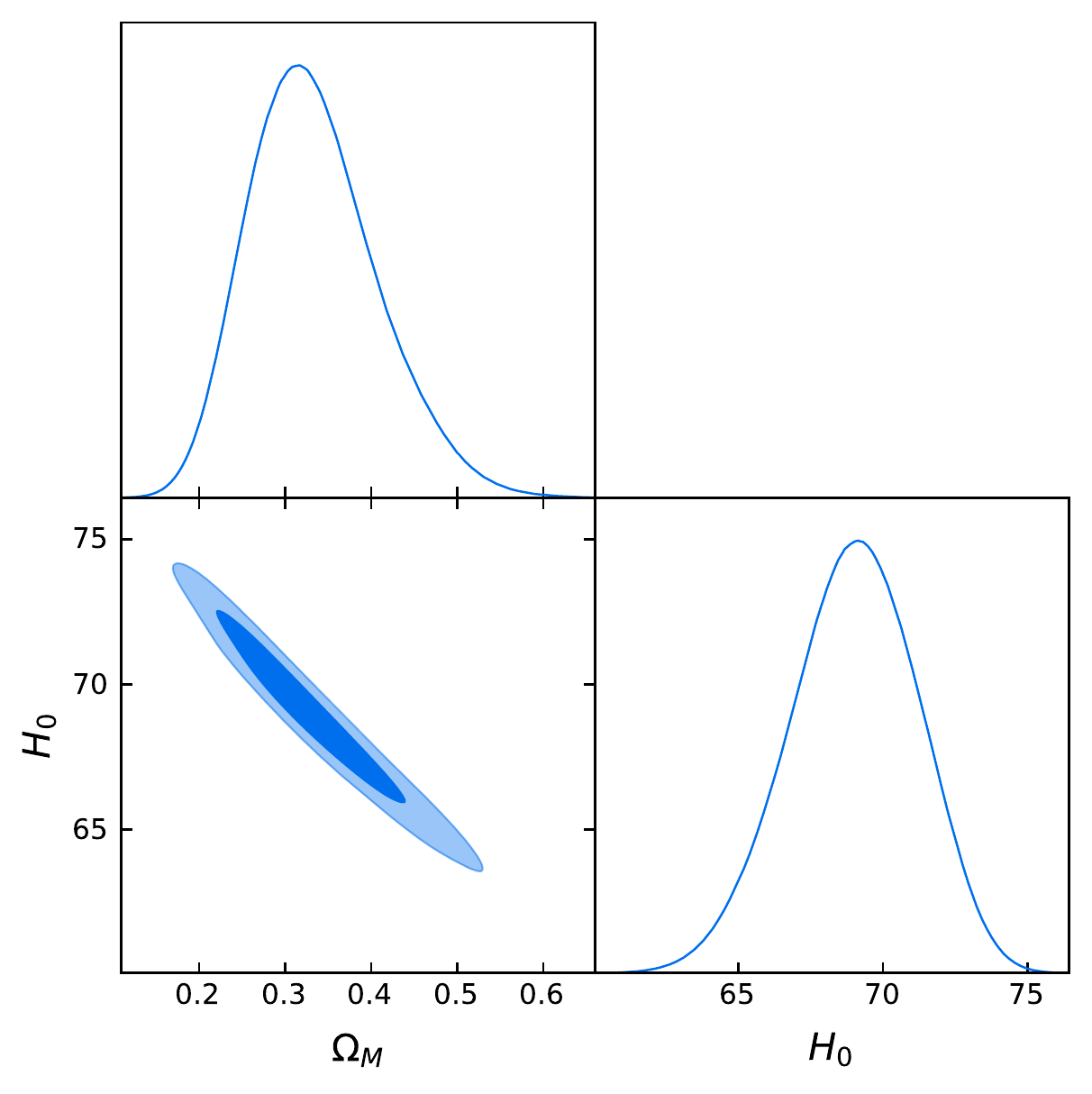}
\includegraphics[width=0.5\hsize,height=0.4\textwidth,angle=0,clip]{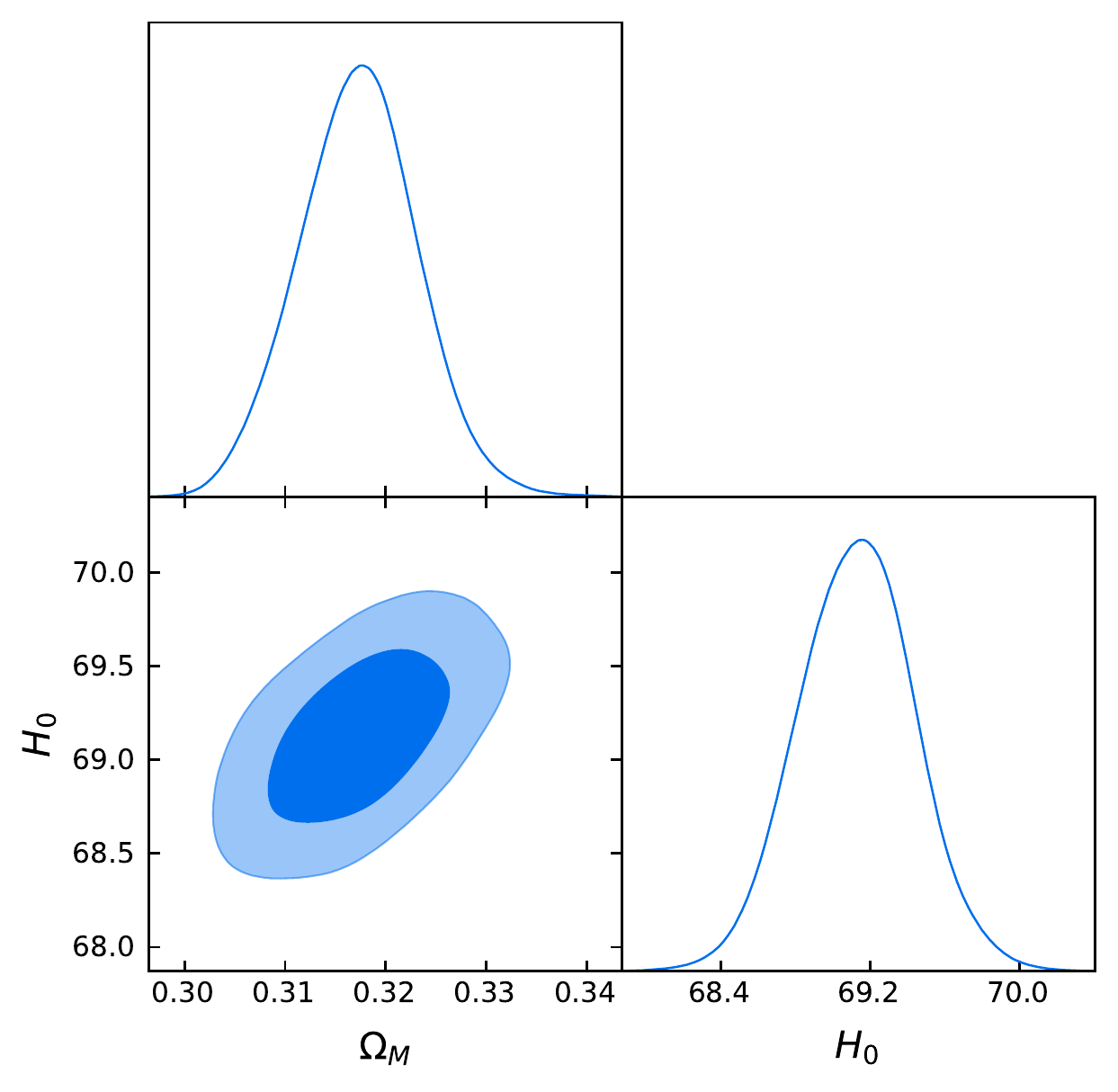}
\includegraphics[width=0.5\hsize,height=0.4\textwidth,angle=0,clip]{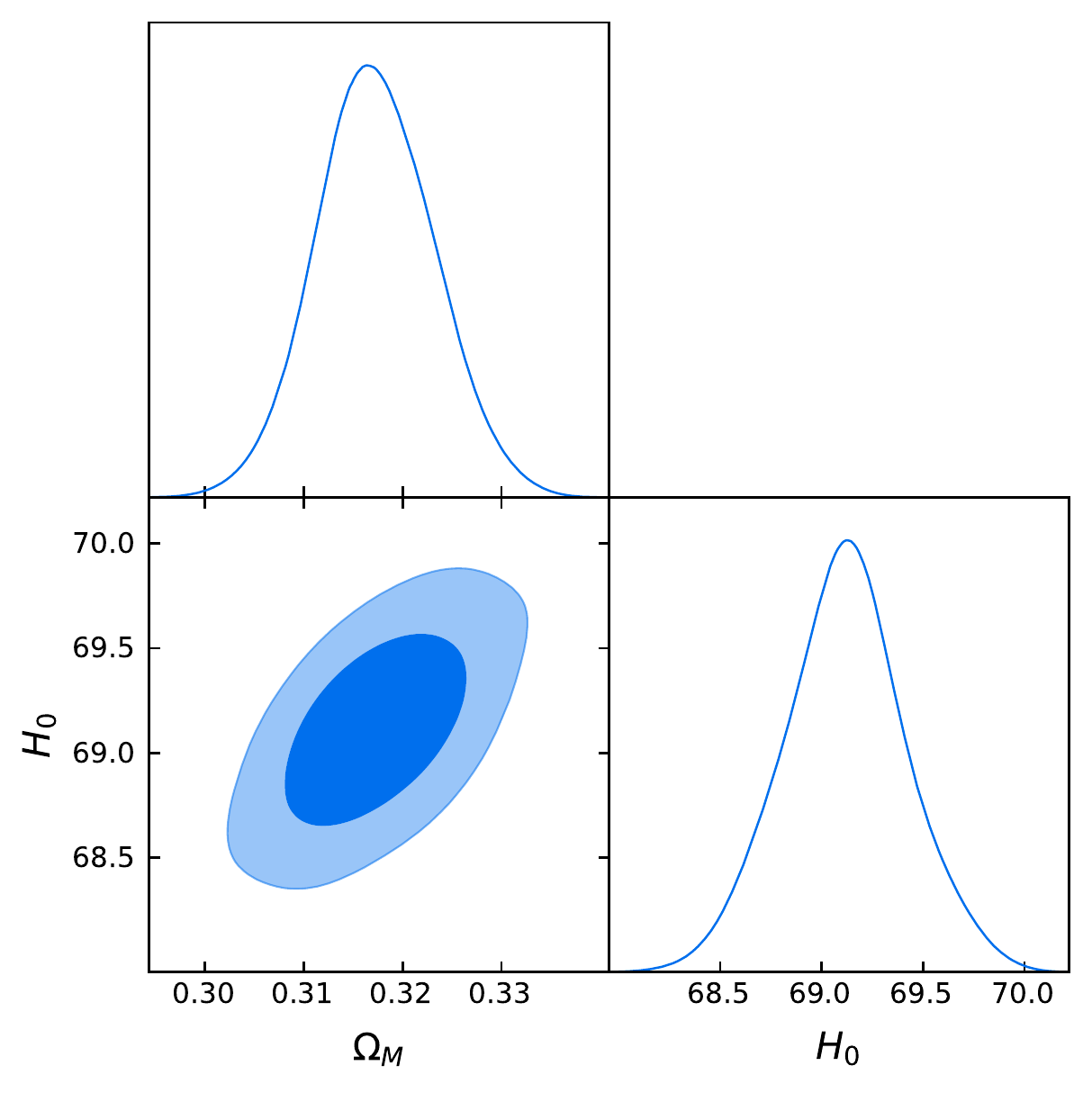}
\includegraphics[width=0.5\hsize,height=0.4\textwidth,angle=0,clip]{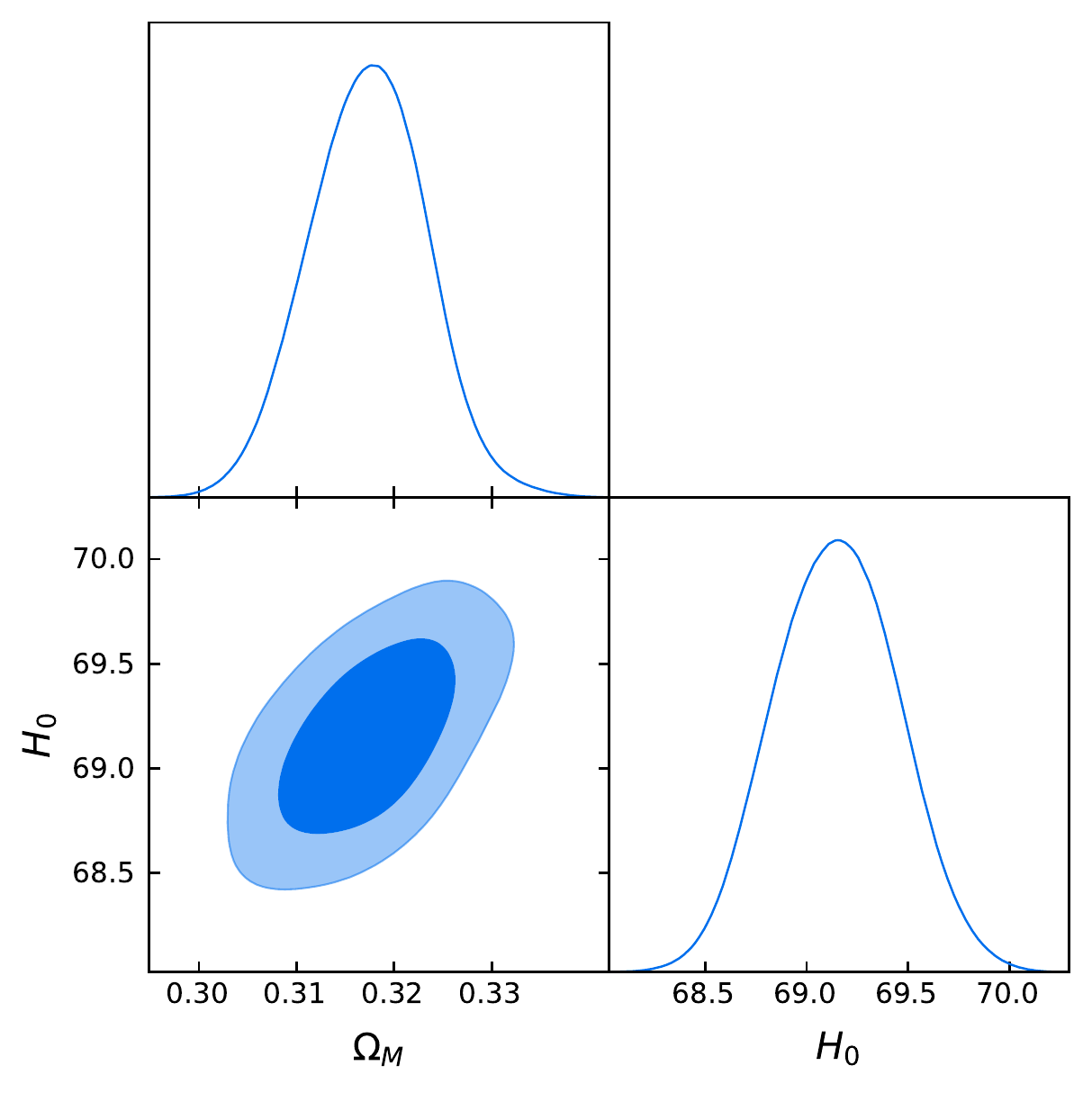}

\caption{\textbf{Results of the cosmological computations of bin 5, varying both $\Omega_M$ and $H_0$, for SNe Ia (top left panel) SNe Ia+BAOs (top right panel), and GRBs+SNe IA+BAOs,  both without (bottom left panel) and with (bottom right panel) considering the evolutionary effects.}}
\label{OMH0BIN5}
\end{figure}

We organize the results as follows: in each table from Table 4 to Table 6 we display the results of a given cosmological parameter. Namely, we show $\Omega_M$ for Table 4 when we fix $H_0$, $H_0$ for Table 5, when we fix $\Omega_M$, and $\Omega_M$ and $H_0$ together for Table 6. We note that the 5 rows of these tables refer to the redshift intervals shown in Table 2. We show the posterior plots for the full sample in figure 4.
We also show the correspondent figures visualizing the SNe Ia, SNe Ia+ BAOs, and SNe Ia+ BAOs+ GRBs posterior plots for bin 5 in figures \ref{OMBIN5}, \ref{H0BIN5}, \ref{OMH0BIN5}. Here we show the posteriors only for the cosmological parameters. These results are obtained using the different bins and the different combination of probes allowed by each bin, and both with and without considering the corrections due to GRBs' evolutionary effects. All the results shown here regarding $H_0$ will be in $km \hspace{1ex} s^{-1} Mpc^{-1}$ units. 


\begin{table}
\centering

\label{table_OM_GRBsSNeBAO5bins}
\begin{tabular}{|c|c|c|c|}
\hline
$\Omega_M$ (SNe Ia) & $\Omega_M$ (SNe Ia+BAOs) & $\Omega_M$ (SNe Ia+BAOs+GRBs) No EV & $\Omega_M$ (SNe Ia+BAOs+GRBs)  EV \\ \hline
$0.387 \pm 0.143$ & - & -& - \\\hline
$0.259 \pm 0.038$ & $0.298 \pm 0.031$ & -& - \\\hline
$0.334 \pm 0.025$ & - & -& - \\\hline
$0.272 \pm 0.016$ & $0.314 \pm 0.006 $ & -& - \\\hline
$0.295 \pm 0.013$ & $0.326 \pm 0.005$ & $0.326 \pm 0.005$ & $0.326 \pm 0.005$ \\\hline
\end{tabular}
\caption{The Table shows in the 1st column the results for $\Omega_M$ for the bins of SNe Ia alone, in the 2nd column the results for $\Omega_M$ with SNe Ia+BAOs, in the 3rd column the same results for SNe Ia+BAOs+GRBs without accounting for the evolutionary effects for GRBs, and in the last column the same combination considering the evolutionary effects.}
\end{table}

\begin{table}
\centering

\label{table_H0_GRBsSNeBAO5bins}
\begin{tabular}{|c|c|c|c|}
\hline
$H_{0}$ (SNe Ia) & $H_{0}$ (SNe Ia+BAOs) & $H_{0}$ (SNe Ia+BAOs+GRBs) No Ev & $H_{0}$ (SNe Ia+BAOs+GRBs) Ev \\
$km \hspace{1ex} s^{-1} Mpc^{-1}$ & $km \hspace{1ex} s^{-1} Mpc^{-1}$ & $km \hspace{1ex} s^{-1} Mpc^{-1}$ & $km \hspace{1ex} s^{-1} Mpc^{-1}$ \\\hline
$70.499 \pm  0.402$ & - & - & - \\\hline
$69.555 \pm  0.292$ & $69.519 \pm 0.315$ & - & - \\\hline
$70.373 \pm 0.347$ & - & - & -\\\hline
$69.456 \pm 0.332$ & $69.159 \pm 0.265 $ & - & - \\\hline
$69.804 \pm 0.450$ & $68.607 \pm 0.261$ & $68.607 \pm 0.262$ & $68.607 \pm 0.261$ \\\hline
\end{tabular}
\caption{Similarly to Table 4, this Table shows the results obtained for $H_0$ for the same bins used previous table fixing the values of $\Omega_M.$} 
\end{table}

\begin{table}
\centering

\label{table_OMH0_GRBsSNeBAO5bins}
\begin{tabular}{|c|c|c|c|}
\hline
 $\Omega_M$ (SNe Ia) & $\Omega_M$ (SNe Ia+BAOs) & $\Omega_M$ (SNe Ia+BAOs+GRBs) No Ev & $\Omega_M$ (SNe Ia+BAOs+GRBs) Ev \\
 $H_0$ (SNe Ia) & $H_{0}$ (SNe Ia+BAO) & $H_{0}$ (SNe Ia+BAOs+GRBs) No Ev & $H_{0}$ (SNe Ia+BAOs+GRBs) Ev \\\hline
NA & - & - & - \\
$71.20 \pm 0.88$ & - & -& - \\\hline
$0.60 \pm 0.20$ & $0.429 \pm 0.072$ & - & - \\
$67.4 \pm 1.4$ & $68.679 \pm 0.605$ & -& - \\\hline
 NA & - & - & -\\
$64.8 \pm 2.2$ & - & -& - \\\hline
$0.339 \pm 0.140$ & $0.311 \pm 0.006 $ & - & - \\
$68.875 \pm 2.380$ & $69.271 \pm 0.275$ & -& - \\\hline
$0.314 \pm 0.065$ & $0.317 \pm 0.006$ & $0.317 \pm 0.006$ & $0.317 \pm 0.006$ \\
$69.373 \pm 2.029$ & $69.109 \pm 0.304$ & $69.110 \pm 0.319$& $69.107 \pm 0.312$ \\\hline
\end{tabular}
\caption{Similarly to the two previous Tables, this Table shows the same analysis, but for the case where $\Omega_M$ and $H_0$ are varied together. We indicate with "NA" (not available) the runs for $\Omega_{M}$ that do not converge in the allocated intervals.}
\end{table}

We now comment the results summarized in the Tables described above. We note that, in general, increasing the number of probes used for the computations of each cosmological parameter does indeed increase the precision of our results and, thus, decrease the scatter on the cosmological parameters. Starting from varying $\Omega_M$ and fixing $H_0$ (see Table 4), we note that the highest precision has been reached for the results given by bin 5 when at least SNe Ia and BAOs are considered: $\Omega_M=0.326 \pm 0.005$, which is also the same result obtained once GRBs are taken into account both accounting or not for evolutionary effects. Adding more probes to the PS does indeed increase the precision on the cosmological calculations, because  bin 5 is the one with more BAOs and GRBs (see Table 2). This is true even if the numbers of BAOs and GRBs are significantly smaller than the ones of SNe Ia  (we stress again that the observations of galaxies and cluster of galaxies linked to the \textcolor{green}{16} BAOs sets are hundred of thousands, see \cite{Alam2021}).  This is the case even after the division in bins: 212 SNe Ia vs {\bf 11} BAO measurements vs 50 GRBs in bin 5, where we find the highest number of {\bf non-SNe Ia} probes. Regarding the central values obtained for each bin, we note that all the results for $\Omega_M$ are consistent within {\bf 1} $\sigma$ once we take into account the BAOs. We also note that BAOs have the most relevant effect concerning the decrease of the cosmological uncertainties. The decrease on the scatter ($\sigma$) of $\Omega_{M}$ for bin 5 compared to the SNe Ia alone is $62\%$ for GRBs+BAO+SNe Ia both for the case of evolution and for the case of no evolution and also for the case of SNe Ia +BAO only.

The results pertinent to $H_0$ when we fix $\Omega_M$ are shown in Table 5. We note that in this case the best precision is reached again by  bin 5 when we consider BAOs+SNe Ia ($H_0=68.607 \pm 0.261$) and BAOs+SNe Ia+GRBs with evolution ($H_0=68.607 \pm 0.261$), followed by BAOs+SNe Ia+GRBs without evolution ($H_0=68.607 \pm 0.262$). We would like to stress again the increased precision on the cosmological results obtained by adding more probes to the SNe Ia. The result is very clear when considering the values obtained by bin 5: we start from $H_0=69.804 \pm 0.450$ for SNe Ia only, to arrive to  $H_0=68.607 \pm 0.261$ for both the SNe Ia+BAOs and
 for SNe Ia+BAOs+GRBs with the evolutionary correction. For the SNe +BAO+ GRBs results, we thus obtain a percentage decrease on the scatter equal to 42\% with respect to the SNe Ia only and for the case of non-evolutionary and evolutionary effects or when we consider BAO+SNe only. 
 

We now focus on the case where we vary $\Omega_M$ and $H_0$ contemporaneously, shown in Table 6. We note that for the bins 1 and 3 when we consider SNe Ia alone, SNe + BAOs and SNe + BAO + GRBs the computations do not converge for the prior interval $0<\Omega_M<1$: these cases are indicated with NA inside the table. 
The scatter is reduced by adding other probes to the SNe Ia, especially due to BAOs. In particular, for  bin 4, when we consider SNe Ia+BAOs, we reach a precision for $\Omega_M$ and $H_0$  similar to the one obtained when we vary these two parameters alone:  $H_0=69.271 \pm 0.275$ when we vary them contemporaneously vs. $H_0=69.159 \pm 0.265$ when we keep $\Omega_M$ fixed, and $\Omega_M=0.311 \pm 0.006$  when we vary both of them vs. $\Omega_M=0.314 \pm 0.006$ when we keep $H_0$ fixed. When we consider bin 5 the same precision is reached for $\Omega_M$ for SNe +BAO, GRBs+SNe +BAO with and without considering evolutionary effects. Regarding instead $H_0$ the most precise value is reached when only BAO+SNe Ia are considered followed by the case with GRBs with evolution. The least precise measurement belong to the case of GRBs with no evolution.
Regarding the percentage decrease on the scatter for the bin 5, for $\Omega_M$ we obtain a $90.8\%$ decrease for all the cases in which we add probes to the SNe Ia, while for $H_0$ we obtain a $85.0\%$ decrease when we compare SNe Ia to SNe Ia+BAOs, a $84.3\%$ decrease when we compare SNe Ia to SNe Ia+BAOs+GRBs without evolution, and a $84.6\%$ decrease when we compare SNe Ia to SNe Ia+BAOs+GRBs with evolution.

\begin{figure}
\includegraphics[width=0.5\hsize,height=0.4\textwidth,angle=0,clip]{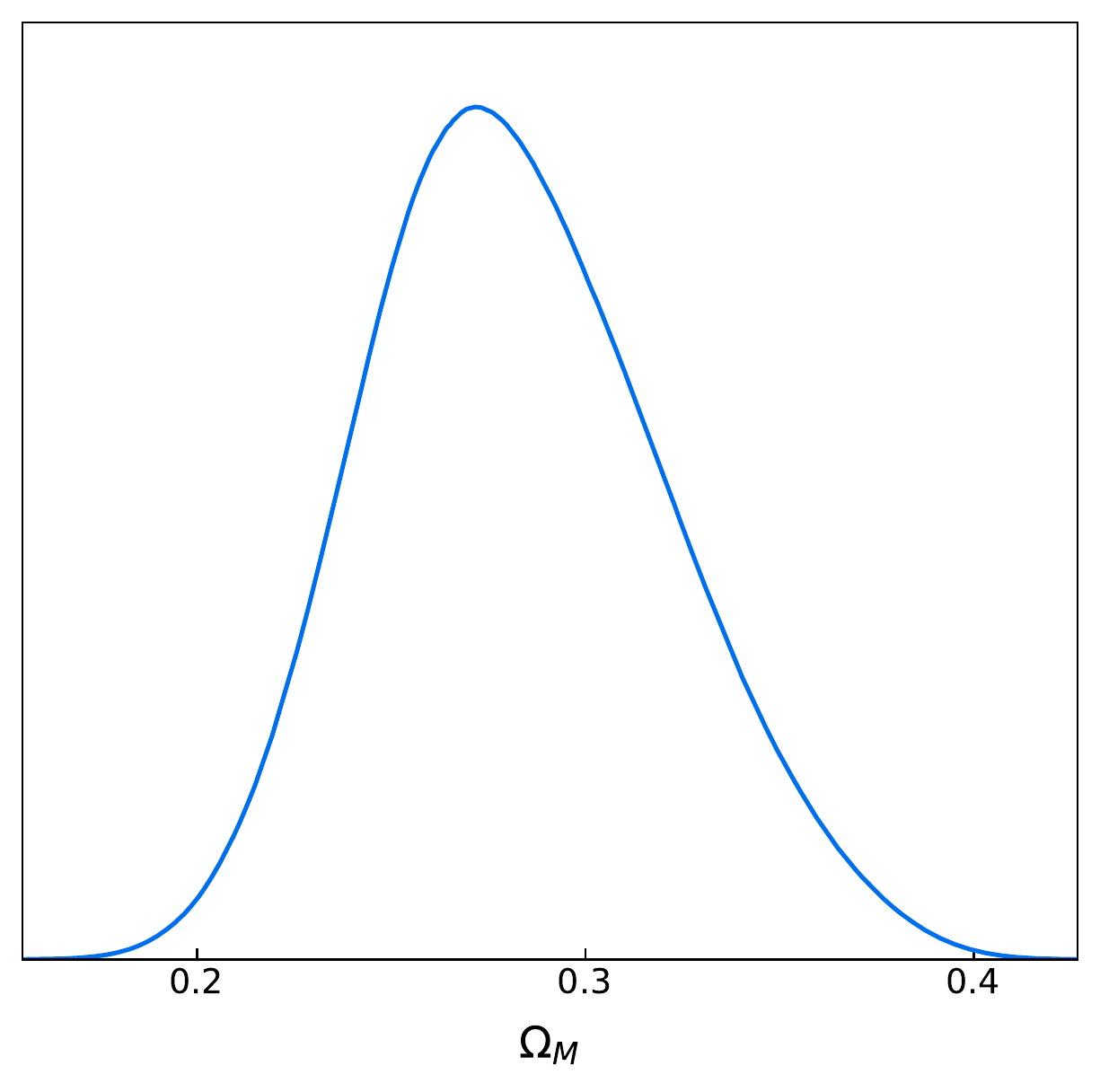}
\includegraphics[width=0.5\hsize,height=0.4\textwidth,angle=0,clip]{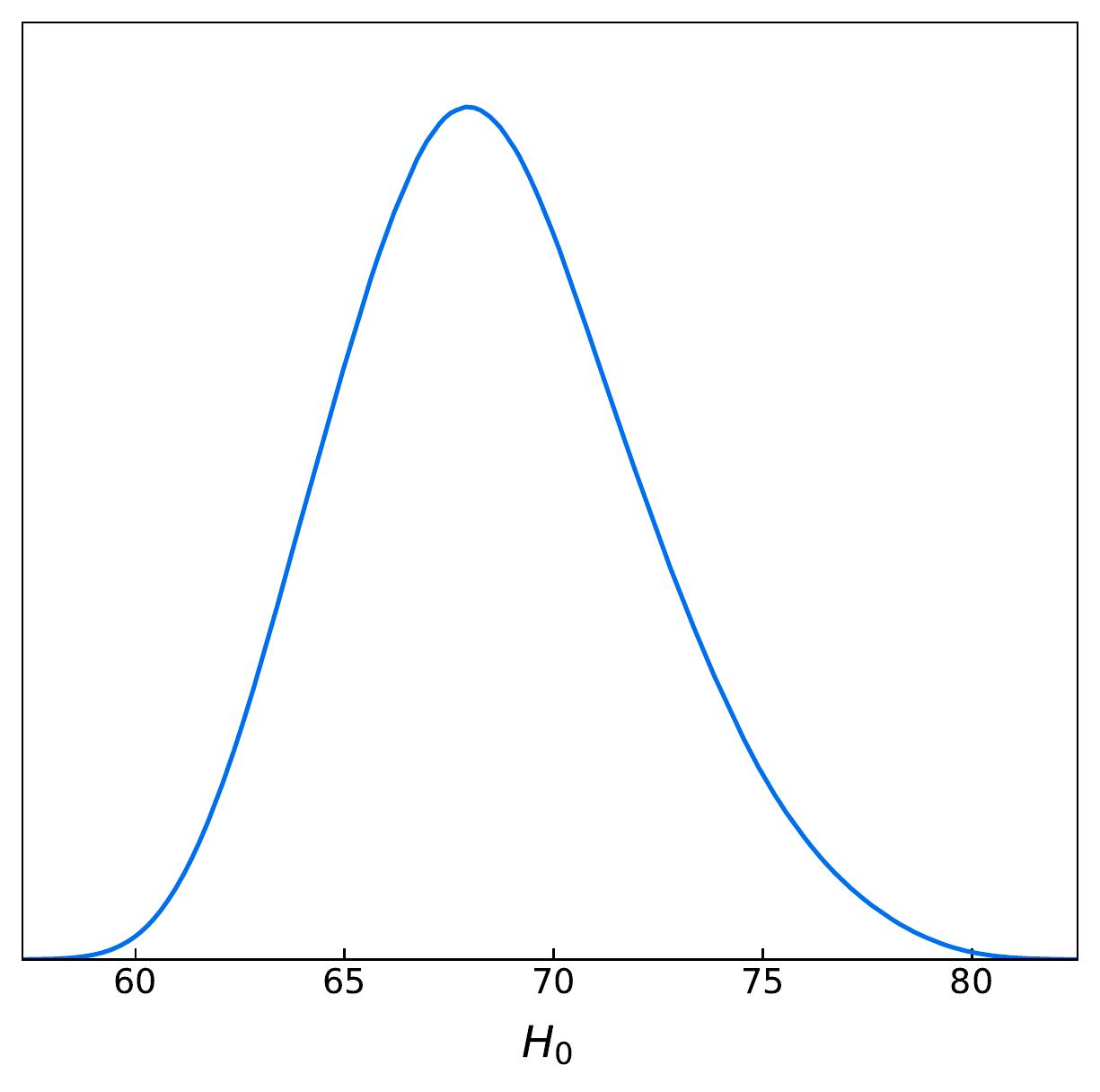}
\includegraphics[width=0.5\hsize,height=0.4\textwidth,angle=0,clip]{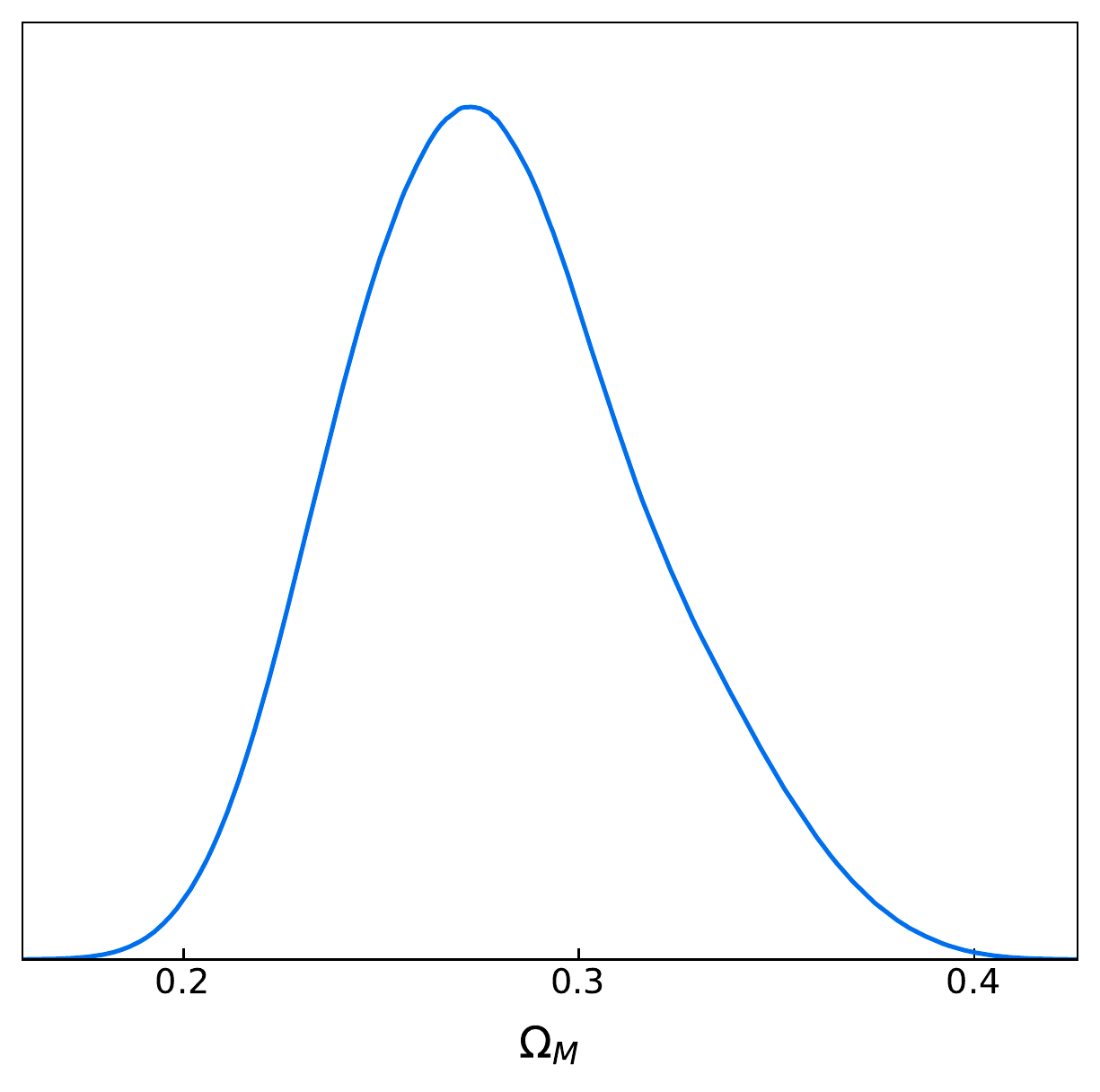}
\includegraphics[width=0.5\hsize,height=0.4\textwidth,angle=0,clip]{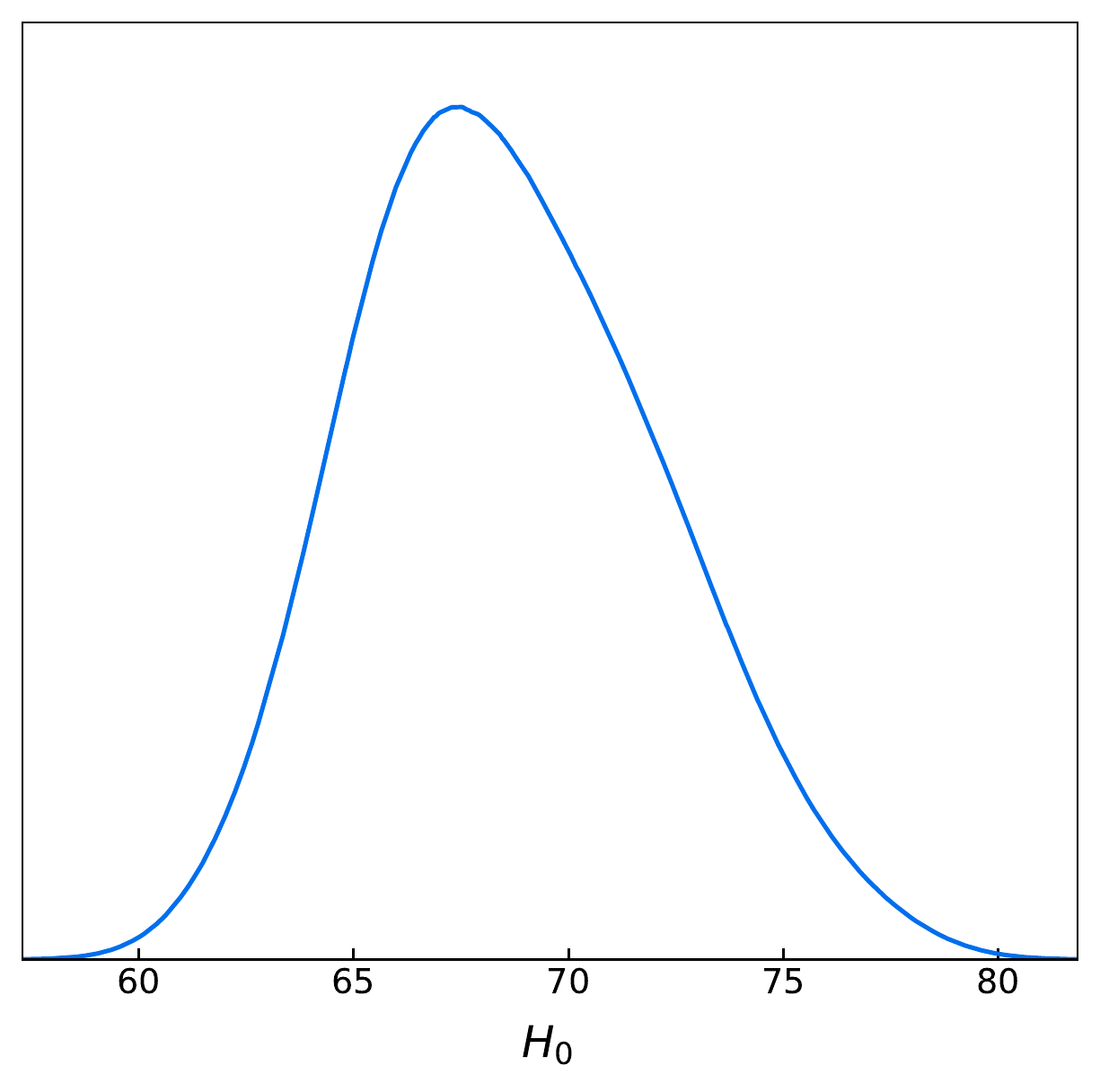}

\caption{\textbf{ Results of the cosmological computations regarding the bin 6, considering $\Omega_M$ alone without evolution (top left panel), $H_0$ alone without evolution (top right panel), $\Omega_M$ alone with evolution (bottom left panel), and $H_0$ alone with evolution (bottom right panel).}}
\label{plotBIN6}
\end{figure}

We now consider a new bin, that we call bin 6, in which we take into account all the probes that go beyond the redshift range of the PS, which we recall is $z=2.26$. The probes that satisfy this requirement are {\bf 2} BAO related measurements and 23 GRBs . These are shown in figure \ref{plotBIN6}and Table 7. We note that, as expected given the smaller sample size, the scatter on the cosmological parameters are larger than the ones obtained for the other bins, but it is still interesting that we obtain closed contours for both $H_0$ and $\Omega_M$ with only these probes. This result has been achieved both with and without considering the evolutionary effects for GRBs.

\begin{table}
\centering

\label{tableBin6}
\begin{tabular}{|c|c|c|c|c|c|}
\hline
$\Omega_M$ (GRBs+BAOs) No Ev & $H_{0}$ (GRBs+BAOs) No Ev & $\Omega_M$ (GRBs+BAOs) Ev & $H_{0}$ (GRBs+BAOs) Ev \\\hline
$0.272 \pm 0.038$ & $67.833 \pm 3.499$  & $0.271 \pm 0.037$ & $68.450 \pm  3.601$\\\hline
\end{tabular}
\caption{Results obtained considering the BAOs and GRBs belonging to bin 6 varying: 1) $\Omega_{M}$ alone without evolution; 2)  $H_0$ alone without evolution; 3) $\Omega_{M}$ alone with evolution, and 4) $H_0$ alone with evolution.}
\end{table}

\section{Discussion and conclusions} \label{Conclusions}
Here we have used the samples of SNe Ia, BAOs, and GRBs to infer cosmological parameters, in particular $\Omega_M$ and $H_0$. Regarding the GRBs, we have used the fundamental plane correlation both with and without taking into account the corrections for redshift evolution and selection biases, using a sample selected by us, called the PLAT sample. After this correction, we obtain one of the smallest intrinsic scatters for any GRB correlation involving plateau features. Our total sample is made up of 1048 SNe Ia, 50 GRBs, and {\bf 16} BAO based measurements. 

First, in Table 1 we present the BAO data set used for our computations, while in Table 2 we show the redshift ranges, the number of probes in each redshift range. 
In Table 3 we consider the full samples at our disposal, without the bin division. Here we note that the best precision is obtained by considering the SNe Ia+BAOs sample, and by the BAOs+SNe Ia+GRBs with and without accounting for the evolutionary effects when we vary the values of $\Omega_M$ compared to the SNe Ia alone and BAOs alone cases.

\begin{figure}
    \centering
    \includegraphics[width=0.6\hsize,height=0.6\textwidth,angle=0,clip]{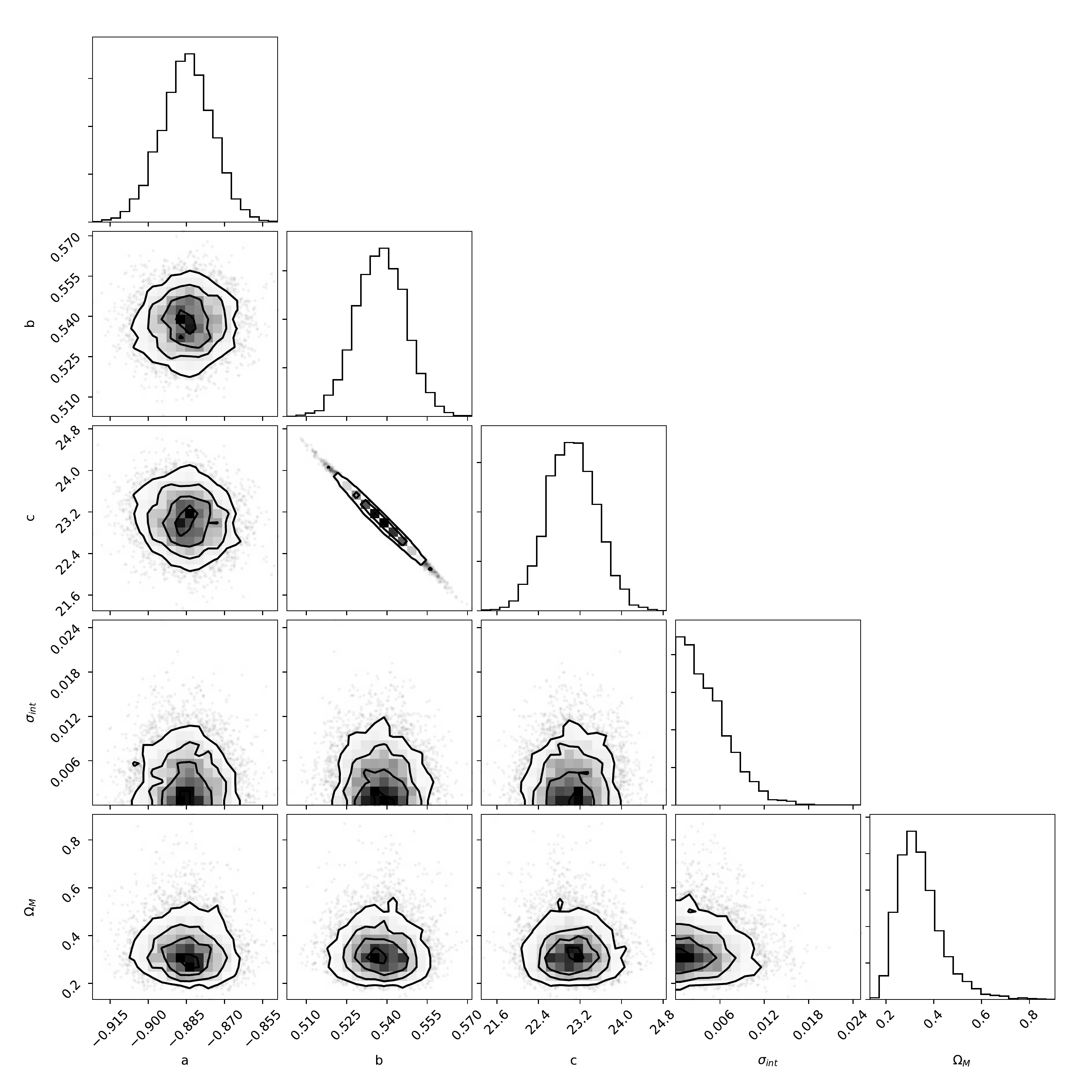}
    \caption{Posterior contours for $\Omega_M$ considering 1000 simulated GRBs on the platinum fundamental plane. }
    \label{simulation}
\end{figure}

We then proceed with the cosmological computations of the data in each bin. Adding more probes to the SNe Ia sample does indeed decrease the uncertainty on the cosmological parameters. 
This statement is true for all the cases analyzed and for all bins. Indeed, a great beneficial effect is due to the inclusion of both the BAO measurements and GRBs. The biggest improvements have been reached when we consider bin 4 and bin 5, where we find more BAO and GRBs. One relevant example for this aspect is represented by the $H_0$ computation for bin 5, where we note a significant decrease in the scatter for the SNe Ia+BAOs+GRBs subset with respect to the SN Ia only: 42\% for the non evolutionary and  the evolutionary cases. Even if we do not see a clear reduction of the scatter for the cosmological computations when we compare SNe Ia+BAOs+GRBs with the SNe Ia+BAOs, the inclusion of GRBs in our analysis is still important for those cases because they allow us to infer cosmological parameters up to $z=5$ keeping a similar precision on them. A possible reason behind the smaller contribution on the cosmological precision given by the GRBs with respect to BAOs and SNe Ia lies in the higher uncertainties of the observed quantities as well as the smaller number of observed GRBs. Indeed, we stress again the GRB set is more than 4 times smaller than the SNe Ia one even after the division in bin, thus its effect is still relevant and expected to increase after gathering more data. In \citep{Dainotti2022d} we have simulated GRBs to understand how many more years of observations both with present and future telescopes are needed in order for the GRBs to be cosmological tools as reliable as SNe Ia. In particular, we found out that we will need 789 GRBs from the PLAT sample to reach the same precision reached by \cite{Conley} for the SNe Ia. Here, in figure \ref{simulation} we show an example of the simulations we have produced, where 1000 GRBs have been generated on the platinum fundamental plane. 
We note that with this number of simulated GRBs we reach closed contours on $\Omega_M$, with an error equal to $0.077$, thus compatible with the precision reached for some bins by the SNe Ia only with the observed data. 

We here discuss the results on $H_0$ considering the full SNe Ia+BAOs+GRBs set with respect to the CMB measurements in comparison to the results obtained using the full PS alone both by fixing $\Omega_M$ or leaving it to vary. In particular, we proceed by calculating if there are discrepancies with the CMB results considering the formula used in \cite{Dainotti2021a}.  We here report it adapted to our purposes: 

\begin{equation}
    x_{i}=\frac{H_{0,i}-H_{0,CMB}}{\sqrt{\sigma_{H_{0,i}}^2+\sigma_{H_{0,CMB}}^2}}
    \label{discrepancies},
\end{equation}
where i=SNe Ia or GRBs+SNe Ia+BAOs and $x_{i}$ is the tension computed via this method. 
We obtain for the Full PS case when we vary only $H_0$: $x_{SNe Ia}=4.97$, while when we compute the same quantity for the full GRBs+SNe Ia+BAOs without considering the evolutionary effects we obtain  $x_{GRBs+SNe Ia+BAOs No EV}=4.14$, thus we note a percentage decrease of  $16.7 \%$. When we instead consider the evolutionary effects for GRBs, we obtain  $x_{GRBs+SNe Ia+BAOs  EV}=4.15$, thus yielding a reduction of  $16.5 \%$.    

Considering the case where we vary both $H_0$ and $\Omega_M$, for the full PS we compute $x_{SNe Ia}=4.43$, while for the GRBs+SNe Ia+BAOs  we obtain $x_{GRBs+SNe Ia+BAOs No EV}=x_{GRBs+SNe Ia+BAOs EV}=4.37$ , thus in these cases we note a percentage decrease of $1.4 \%$.  

We now take into account the results related to bin 5. When we vary $H_0$ alone we obtain $x_{SNe Ia}=3.57$ for the SNe Ia inside bin 5, while for the GRBs+SNe Ia+BAOs without considering the evolutionary effects we obtain $x_{GRBs+SNe Ia+BAOs No EV}=2.14$, thus we note a percentage decrease of $40.0 \%$. For the case where we consider the evolution, we obtain the same result. This consistent decrease of the scatter is due to the decrease of the central value of $H_0$, that becomes closer to the one obtained by Planck once more cosmological probes are taken into account.

Instead, when we let $\Omega_M$ vary with $H_0$, we compare the results of bin 5 and we note that the discrepancy significantly increases due to the very high scatter on the SNe Ia results ($H_0=69.373 \pm 2.029$). This is $536 \%$ larger than the scatter computed for the GRBs+SNe Ia+BAOs bin 5 without considering the evolutionary effects and 550\% larger if we do consider GRBs corrected for evolutionary effects. Thus, due to the large errors in this case the $H_0$ value for the SNe Ia is compatible with the value of the CMB. 


In summary, our study regarding the cosmological parameters using different probes allows us to conclude that 1) combining more probes is beneficial for the precision of the results, 2) BAOs have a strong relative weight on the computations, 3) GRBs can be added to the cosmological computations carrying similar uncertainties when combined with other probes. This allows us to extend the cosmological ladder to higher redshifts. Moreover, in all cases, when we consider $\Omega_M$ and $H_0$ fixing the other parameter, the addition of GRBs increases or equals the precision on the results obtained without GRBs. We expect that increasing the GRBs sample in a near future will allow us to explore the nature of the cosmological parameters at larger redshifts, given the encouraging conclusions reached by our work. 

\section*{Funding}

M.G.D. acknowledges support from NAOJ and from the Department of Energy of State who funded the summer internship of Srinivasaragavan, R. Wagner, L. Bowden, and R. Waynne.

\section*{acknowledgement}
This paper is based on the data provided by the UK Swift Science Data Centre at the University of Leicester. We are grateful to S. Savastano for the very help in writing of the Python codes used for the analysis, to B. De Simone, T. Schiavone, and S. Nagataki for helping in the SNe Ia analysis,to L. Dash for helping in running the cosmological notebooks, and to V. Nielson for producing the simulations regarding GRBs. We thank again B. De Simone for helping in the writing of the paper related to the SNe Ia and BAO. We are grateful to G. Srinivasaragavan, Z. Nguyen, R. Wagner, L. Bowden, and R. Waynne for helping with the lightcurve parameter fitting. 

\end{document}